\documentclass[aps,amsmath,amssymb,mathabxsuperscriptaddress,showpacs,floatfix,twocolumn, pra,reprint]{revtex4-1}
\usepackage{amssymb,amsthm,latexsym,amsmath,dsfont,pstricks,hyperref,multirow}
\usepackage{graphicx,epsfig,float}
\usepackage{epstopdf}
\usepackage{psfrag}
\usepackage{color} 
\usepackage{subfig}
\usepackage[T1]{fontenc}
\usepackage{natbib}
\usepackage{etoolbox}
\usepackage{booktabs}
\apptocmd{\sloppy}{\hbadness 10000\relax}{}{}

\newcommand{\I}{{\rm{i}}}
\newcommand{\D}{{\rm{d}}}
\newcommand{\E}{e}
\newcommand{\ket}[1]{| #1 \rangle}
\newcommand{\bra}[1]{\langle #1 |}
\newcommand{\braket}[2]{\langle #1 | #2 \rangle}
\newcommand{\ketbra}[2]{| #1 \rangle \langle #2 |}
\newcommand{\onehalf}{{\frac{1}{2}}}

\newcommand{\ie}{i.e.}

\newcommand{\mmin}{{\rm min}}

\newcommand{\RHS}{right-hand side\;\,}
\newcommand{\LHS}{left-hand side\;\,}

\def\real{{\mathbb{R}}}

\def\complex{{\mathbb{C}}}

\def\proba{{\rm I\kern -.18em P}}

\newcommand{\Span}{\operatorname{span}}

\newcommand{\supp}{\operatorname{supp}}

\newcommand{\identity}{{\mathds{1}}}

\newcommand{\tr}{\operatorname{tr}}

\newcommand{\rank}{\operatorname{rank}}
\newcommand{\cotan}{{\rm{cotan}}}

\newcommand{\be}{\begin{equation}}
\newcommand{\ee}{\end{equation}}
\newcommand{\ba}{\begin{eqnarray}}
\newcommand{\ea}{\end{eqnarray}}
\newcommand{\nn}{\nonumber}
\newcommand{\dss}{\displaystyle}

\newcommand{\eps}{\varepsilon}

\newcommand{\rv}{{\bf{r}}}
\newcommand{\sv}{{\bf{s}}}

\newcommand{\xv}{{\bf{x}}}
\newcommand{\yv}{{\bf{y}}}

\newcommand{\sigmav}{{\mathbf{\sigma}}}

\newcommand{\Ff}{{\cal F}}

\newcommand{\Hh}{{\cal H}}

\newcommand{\Kk}{{\cal K}}

\newcommand{\Mm}{{\cal M}}

\newcommand{\Pp}{{\cal P}}

\newcommand{\reg}{{\rm reg}}

\newcommand{\states}{{\cal E}_{\cal H}}
\newcommand{\statesof}[1]{{\cal E}_{#1} }
\newcommand{\geo}{{\mathrm{g}}}
\newcommand{\inv}{{\,\mathrm{inv}}}

\newcommand{\regrhoone}{\rho_1 ( \varepsilon_1)}
\newcommand{\regrhotwo}{\rho_2 ( \varepsilon_2)}
\newcommand{\veceps}{{\mathbf{\varepsilon}}}
\newcommand{\TR}{{\mathrm{TR}}}
\newcommand{\qubit}{{\mathrm{qubit}}}
\newcommand{\FS}{{\mathrm{FS}}}
\newcommand{\Bures}{{\mathrm{Bu}}}
\newcommand{\hor}{{\mathrm{h}}}
\newcommand{\verti}{{\mathrm{v}}}

\begin{document}

\title{Bures geodesics for  non-faithful states and quantum speed limit}

\author{S. Carrasco, D. Spehner}
\affiliation{CI$^2$MA and Departamento de Ingenier\'{\i}a Matem\'atica, Universidad de Concepci\'on, Concepci\'on, Chile}

 \date{\today}

 \begin{abstract}
 The quantum speed limit establishes a bound
 on the minimal time required for a quantum system to evolve from a given initial state to a final state. When the mean energy variance is fixed this limitation is captured by the Mandelstam--Tamm bound. The fastest quantum evolution saturating this bound follows a geodesic arc connecting the two states. Such geodesics  in the manifold of quantum states are explicitly known when the  states are pure (Fubini-Study geodesics) and when they are mixed and given by faithful density matrices (Bures geodesics).  
 In this article we obtain the explicit form of the Bures geodesic arcs joining two non-faithful density matrices,  which may have different ranks. 
 For pure states one recovers the Fubini-Study geodesics.
 A necessary and sufficient condition for the uniqueness of the shortest geodesic arc is given.
When the condition is not fulfilled there are infinitely many such arcs, all having length equal to the arccos Bures distance between the two states, in analogy with the arcs of great circles connecting the two poles of a sphere.    
We discuss the implications of our results for the quantum speed limit.
\end{abstract}


\maketitle

 \section{Introduction.} \label{sec-Intro}

 An important issue for designing faster and optimized information-processing devices is related to the constraint imposed by quantum mechanics on the minimal time needed by a quantum system to evolve between
	two distinct states, known as the quantum speed limit (QSL)~\cite{Deffner_review_QSL}. 
For pure states, two  bounds capture this constraint: the Mandelstam--Tamm (MT)
	bound, expressed in terms of the energy variance~\cite{MandelstamTamm1945},
	and the Margolus--Levitin bound, governed by the mean energy above
	the ground state~\cite{MargolusLevitin1998}. 
	In the geometric reformulation of  Anandan and Aharonov~\cite{AharonovAnandan1990}, the  MT bound is derived by noting that 
	curves representing quantum evolutions in the projective space of pure states have lengths bounded from below by the Fubini-Study distance between the initial and final states. This shows that the bound is saturated when the evolution curve follows a Fubini-Study geodesic.
	As shown by Uhlmann~\cite{Uhlmann91}, this geometric approach extends naturally to mixed states by considering the Bures metric (also known as the Quantum Fisher Information (QFI) metric) on the manifold of density matrices. 
	A MT-like bound for general quantum processes has been derived in~\cite{Taddei2013} by bounding from below the integral of the square root of the QFI along an evolution path by the Bures arccos distance between the initial and final states. Similar bounds based on other monotone metrics have been considered in~\cite{Adesso2016}. As in the pure state case, these bounds are
	saturated when the evolution follows a geodesic. 
	
	The geometry of the manifold of quantum mixed states is more intricate than that of the projective space of pure states, in particular due to the stratification of its boundary into sets of density operators with fixed ranks~\cite{Dittmann1995,Grabowski2005,Anderson_PhD_thesis}. 
	The Bures geodesics on the open manifold of faithful states (density operators with full rank) have been determined in~\cite{Ericson05,Barnum_thesis,moi_geodesics_QMetrology}. The results of these references 
	extend to geodesic arcs joining faithful and non-faithful states. However, determining geodesic arcs joining two density operators which are not of full rank is an open problem.

	Apart from being an interesting geometrical problem on its own, the study of such geodesics also has important applications. As stressed above,  geodesic arcs give the fastest possible quantum evolutions to transform a given state into another state under the constraint of a mean fixed energy variance. Such fastest evolutions have been observed recently using  a superconducting device~\cite{experiment_QSL}.
	Therefore, geodesics offer a route to design faster information-processing devices in quantum technology,  and provide insight into fundamental issues in quantum thermodynamics~\cite{DeffnerLutz2010ThermodynamicLength,DeffnerLutz2013ThermodynamicLength,Scandi2019ThermodynamicLength}. 
	Furthermore, geodesic evolutions lead to the highest precision in quantum metrology for  quantum systems coupled to an ancilla when measurements on the ancilla are not possible~\cite{moi_geodesics_QMetrology} and  can be used to design efficient algorithms for finding  close-to-optimal control parameters in incoherent quantum control~\cite{Mauro-thesis,Giovannetti2009,Deffner2014}. 
	
In this paper we determine the shortest geodesic arcs connecting two non-faithful states.
We show that such arcs are not always unique. Non-uniqueness may occur even when neither the supports nor the kernels of the two states are orthogonal. In such cases there are infinitely many shortest geodesic arcs joining the two states, all having length equal to the Bures arccos distance between the two states, yielding to a family of dynamical evolutions saturating  the MT bound. For two pure orthogonal states, this family consists of infinitely many Fubini--Study geodesics, corresponding to pure state evolutions, together with infinitely many Bures geodesics of rank two having support in the span of the two states.

 The paper is organized as follows. In Sec.\ref{sec-bulk_and_boundary_geodesics} we discuss the special case of a qubit, recall results from the literature on geodesics joining faithful states and give an overview of our subsequent results. We review the geometric QSL and discuss the application of geodesics    
as the fastest quantum evolution in Sec.~\ref{sec_QSL}. In Secs.~\ref{sec-regularization_method} and~\ref{sec-extension_method} we present two alternative methods to tackle the problem of  determining the geodesics when the starting and ending states are not faithful. Our general results are illustrated for qutrits in Sec.~\ref{sec-qutrit}
and for two-qubit systems in Sec.~\ref{sec-2qubit}. The last section~\ref{sec-conclusion} presents our conclusions and perspectives. Technical proofs and numerical results are given in the three appendices.

\section{Bulk and boundary geodesics} \label{sec-bulk_and_boundary_geodesics}
\subsection{Geodesics for a qubit} \label{sec-geodesics_qubit}


\begin{figure*}[t]
	\centering
	\begin{minipage}{0.3\textwidth}
		\centering
		\includegraphics[width=\linewidth]{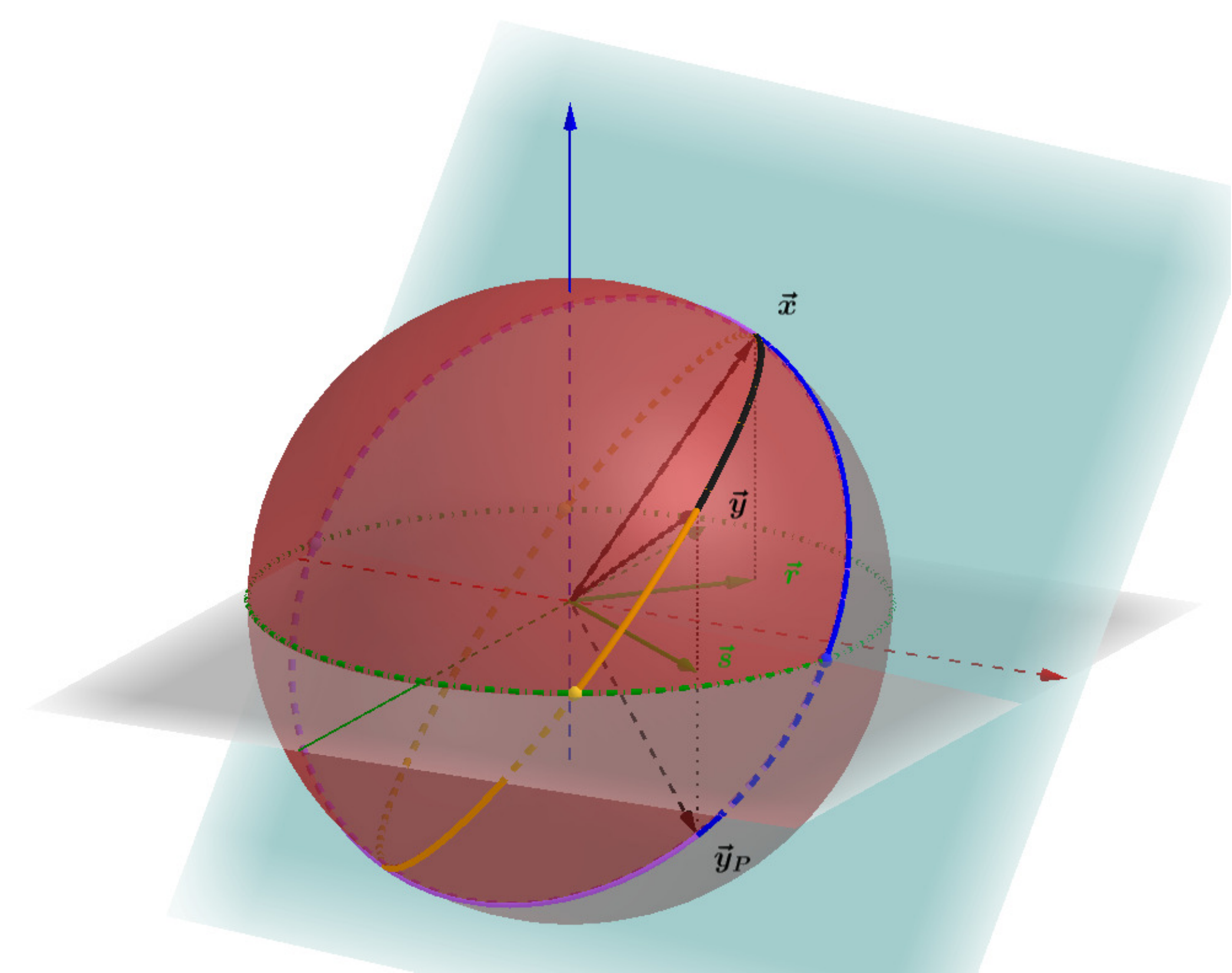}
		
		\vspace*{3mm}
		\includegraphics[width=5cm]{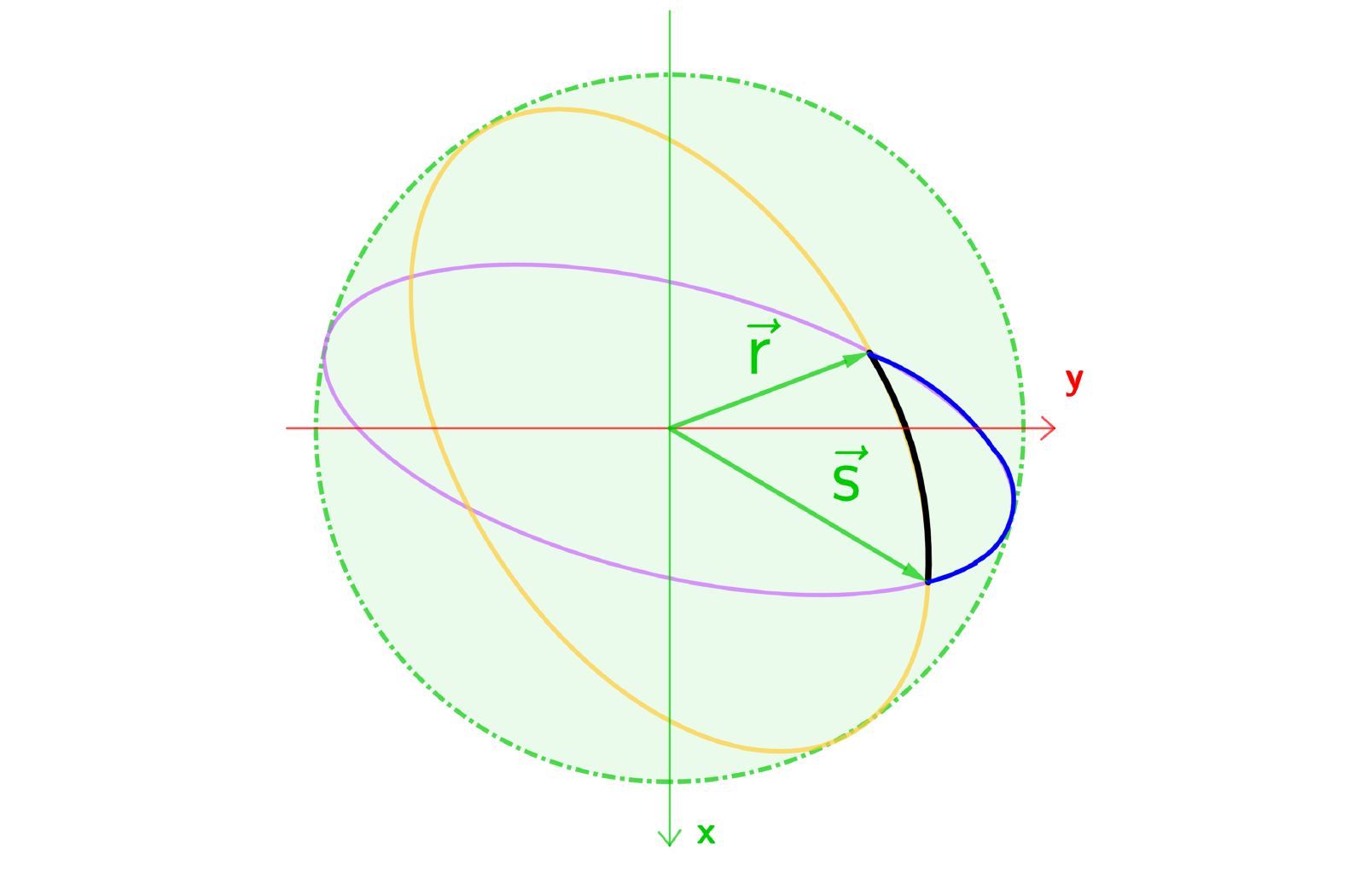}

	\text{(a)}	
	\end{minipage}
	\hspace{2mm} 
	\begin{minipage}{0.3\textwidth}
		\centering
		\hspace*{5mm} \includegraphics[width=\linewidth]{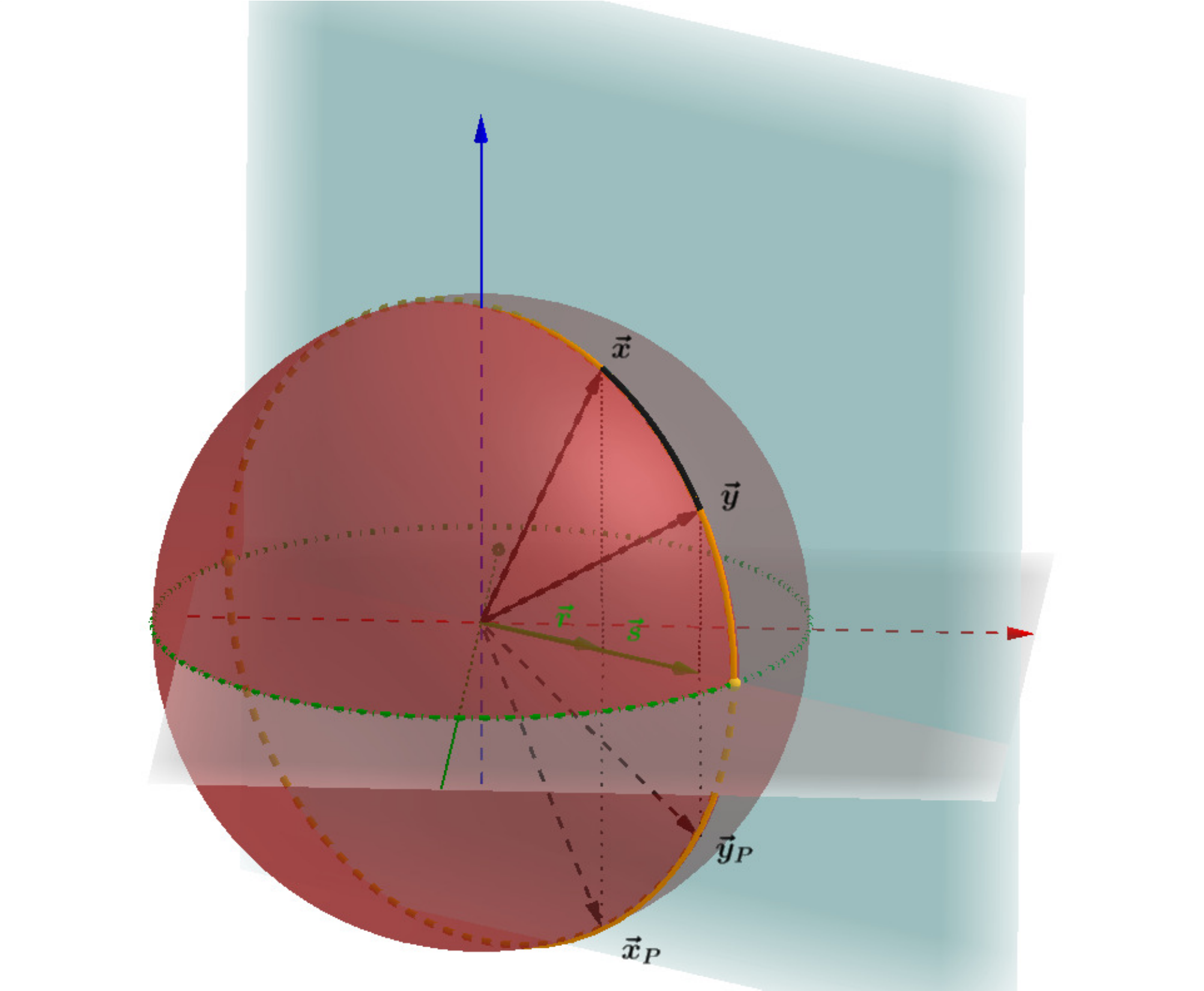}
		
		\vspace*{3mm}
		\includegraphics[width=4.5cm]{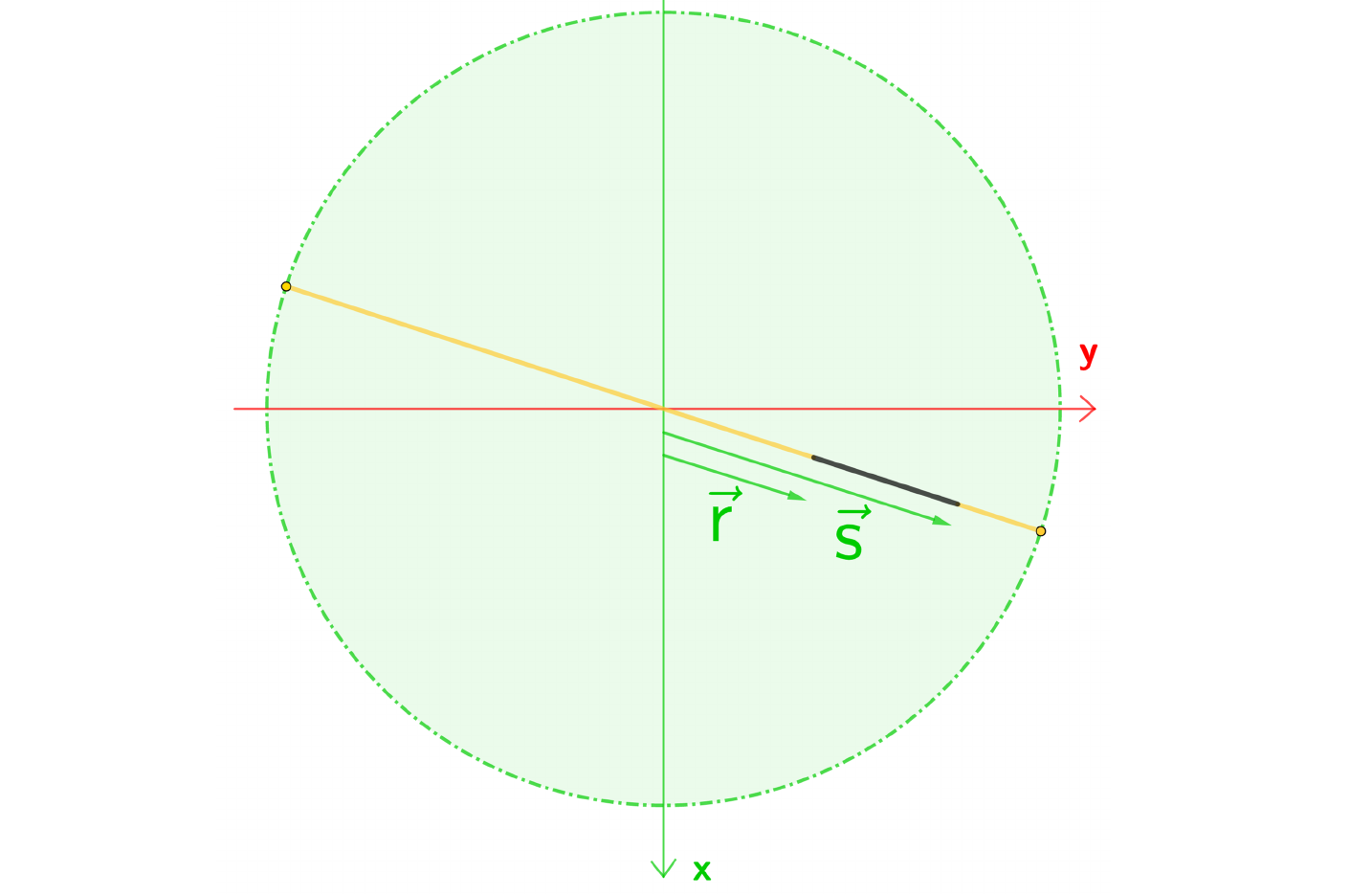}
		
			\text{(b)}	
	\end{minipage}
	\begin{minipage}{0.3\textwidth}
		\centering
		\includegraphics[width=4cm]{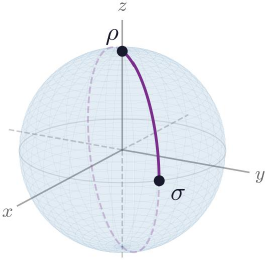}
		
		\vspace{3mm}
		\includegraphics[width=4cm]{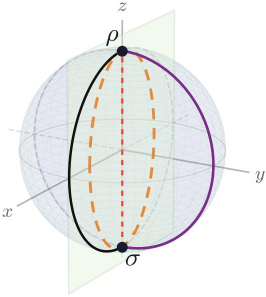}
		
			\text{(c)}
	\end{minipage}
	\captionsetup{format=plain,justification=raggedright}
	\caption{{\textbf{(a)}} Geodesic arcs connecting two faithful non-commuting qubit states, 
		represented on the hypersphere $S^3$ {\textbf{(top)}} and in the section of the Bloch ball spanned by the two Bloch vectors $\rv$ and $\sv$ {\textbf{(bottom)}}.
		 (The sphere on the top represents a 3-dimensional sphere $S^3$, so that its intersection with the horizontal plane (green circle) corresponds to the 2-dimensional Bloch sphere).
		There are four arcs  joining $\rv$ and $\sv$, given by the projections  onto the horizontal
		hyperplane of the  black and orange arcs of the great circle of $S^3$ passing through $\xv$ and $\yv$ and of the blue and violet arcs of the  great circle passing through
		$\xv$ and $\yv_P$.
		{\textbf{(b)}} Same for two commuting faithful states with Bloch vectors $\rv$ and $\sv= \alpha\,\rv$ with $\alpha \not= 1,-1$. The geodesics are the black and orange segments of a diameter of the Bloch ball.
		{\textbf{(c)}}~Geodesics   in the Bloch ball connecting two  pure qubit states $\rho=\ketbra{\psi}{\psi}$ and $\sigma=\ketbra{\phi}{\phi}$.
		 \textbf{Top}: If $\ket{\psi}$ and $\ket{\phi}$ are not orthogonal,
		 there is a unique shortest geodesic (thick violet  solid arc).
		 The complementary great-circle arc is shown as a thin dashed curve.	
		 \textbf{Bottom}: If $\ket{\psi} \perp \ket{\phi}$, there are infinitely many geodesics of equal
		 length connecting $\rho$ and $\sigma$.
		 These include Fubini--Study geodesics  on the Bloch sphere (thick black and violet meridians) as well as ellipses (dashed orange line) and a straight
		 diameter (thin red dashed line)  in the interior of the Bloch ball.	
	}
	\label{fig-geodesics_qubit}
\end{figure*}


We first focus on the qubit case, for which the Bures geodesics connecting faithful as well as non-faithful states can be easily determined.
Let $\statesof{\qubit}$ be the set of all qubit states, \ie, all density operators $\rho: \complex^2 \to \complex^2$ with $\rho \geq 0$ and $\tr \rho=1$, equipped with the Bures distance (see Eq. (\ref{eq-Bures_arccos_dist}) below).
It is known that  $\statesof{\qubit}$ can be identified with a half hypersphere $S_+^3 \subset \real^4$ of radius one and dimension three~\cite{Zuckowski_book}. In fact, in the specific case of a qubit, the Bures metric coincides with the usual metric on $S^3$ up to a factor of  one fourth.
The projection  onto the horizontal hyperplane $\Pp : \xv \in \real^4 \mapsto \rv  \in \real^3$  maps in a one-to-one way any point $\xv \in S_+^3 $ with Cartesian coordinates 
 $(x,y,z,\sqrt{1-x^2-y^2-z^2})$
 to a Bloch vector $\rv=(x,y,z)$,  $| \rv| \leq 1$, hence 
 defining a global coordinate system on $\statesof{\qubit} \sim S_+^3$. The Bloch vector $\rv$ 
 is related to the qubit density operator $\rho$ by $\rho = ( \identity + \rv \cdot \sigmav)/2$, where $\sigmav =(\sigma_1,\sigma_2,\sigma_2)$ is the Pauli matrix vector.
 A density operator $\rho$ is faithful if and only if $\rv$ is inside the Bloch ball, and is non-faithful if and only if it is a pure state $\rho = \ketbra{\psi}{\psi}$ lying on the Bloch sphere $| \rv | = 1$.

The qubit geodesics, being geodesics on the half hypersphere $S^3_+$, are given by arcs of great circles on $S^3$, $\xv_\geo (\tau) = (\sin (2 \theta-\tau)\, \xv + \sin (\tau) \, \yv )/\sin (2 \theta)$,
where $\xv$ and $\yv \in S^3_+$ are the starting and ending points and $\cos (2 \theta) =  \xv \cdot \yv$
(we assume that $\xv \not= \pm \yv$, so that $\theta \not= 0,\pi/2$). 
Due to the factor of one fourth between the Bures and unit hypersphere metrics, to obtain geodesics with unit velocity one must change time as $\tau \rightarrow 2 \tau$. Thus in the Bloch coordinates
 the geodesic arc starting at $\rv= \Pp \,\xv$ and ending at $\sv = \Pp \, \yv$ reads $ \rv_\geo (\tau) = \Pp\, \xv_\geo (2 \tau) $, \ie,
 \begin{equation} \label{eq-Bures_geodesic_qubit_states}
 \rv_\geo (\tau) = 
 \frac{1}{\sin ( 2 \theta)} \Big( \sin ( 2 \theta - 2 \tau)\, \rv + \sin ( 2 \tau)\, \sv \Big)\,,
 \end{equation}    
$0 \leq \tau \leq \theta$, where 
the geodesic length $\theta$ is given by
 \begin{equation}\label{eq-fidelity_for_qubit_states}
 \cos^2 \theta = \frac{1 + \xv \cdot \yv}{2}  
 = \onehalf \Big( 1 + \rv \cdot \sv + \sqrt{(1 - |\rv|^2)(1-|\sv|^2)} \Big)\;.
 \end{equation}

\vspace{2mm}

\noindent {\it Bulk geodesics.}

Let us first discuss geodesics joining faithful states $\rho$ and $\sigma$, having Bloch vectors $\rv$ and $\sv$  inside the Bloch ball, $|\rv|,|\sv| <1$. 
We call such curves {\textit{bulk geodesics}}.
As shown in Fig.~\ref{fig-geodesics_qubit}(a), if 
$\rv$ and $\sv$ are not proportional, \ie, if $\rho$ and $\sigma$ do not commute,
the geodesics passing through $\rv$ and $\sv$  are ellipses centered at the origin $O$ contained in the plane spanned by $\rv$ and $\sv$,  with unit major semi-axis length. 
There are two such ellipses, obtained  by projecting the two great circles of $S^3$ passing through $\xv$ and $\yv$ and through $\xv$ and $\yv_P$, where $\yv_P$ is the symmetric of $\yv$ under the reflexion w.r.t. the horizontal hyperplane (note that $\Pp \, \yv_P = \Pp \, \yv = \sv$). Both ellipses are given by (\ref{eq-Bures_geodesic_qubit_states}), with $\theta$ as in (\ref{eq-fidelity_for_qubit_states}) for the first one and $\theta$ replaced by $\theta_P = \arccos \sqrt{(1 + \xv \cdot \yv_P)/2} >\theta$ for the second one.     
They form four geodesic arcs joining $\rv$ and $\sv$, having lengths $\theta$, $\theta_P$, $\pi - \theta$ and $\pi-\theta_P$. 
Only one of these arcs connects $\rv$ and $\sv$ without intersecting the boundary (Bloch sphere of pure states). This corresponds to the  shortest geodesic, having length $\theta$. It is easy to see that the projections of the great circles
joining $\xv_P$ and $\yv$ and joining $\xv_P$ and $\yv_P$ do not give rise to new geodesics.

If the states $\rho$ and $\sigma$ commute, their Bloch vectors $\rv$ and $\sv$ are proportional and the two aforementioned great circles coincide. As shown in  Fig.~\ref{fig-geodesics_qubit}(b), assuming that $\rv \not= - \sv$ there are only two geodesic arcs joining $\rv$ and $\sv$, which are contained in a diameter of the Bloch sphere. Such straight segments describe paths of density matrices commuting with $\rho$ and $\sigma$ at all times. 
In contrast, if $\rv=-\sv$ then $\xv$ and $\yv_P$ are diametrically opposite on $S^3$, thus there are infinitely many great circles joining them, leading to infinitely many elliptic arcs connecting $\rv$ and $\sv$.
The shortest geodesic arc is nevertheless unique and given by the straight segment joining $\rv$ and $\sv$, as in the case $\rv = \alpha \,\sv$, $\alpha \not= -1$.   
The case $\alpha=-1$ corresponds to commuting states having equal eigenvalues and interchanged eigenvectors.

\vspace{2mm}

\noindent {\it Geodesics connecting pure states.}

Let us now consider two pure states $\rho= \ketbra{\psi}{\psi}$ and $\sigma  =\ketbra{\phi}{\phi}$ with  Bloch vectors $\rv$ and $\sv$ on the Bloch sphere. 
The corresponding vectors $\xv$ and $\yv \in S_+^3$ are contained in the horizontal hyperplane, so that $\rv = \xv$ and $\sv=\yv$. Since the great circles of $S^2$ are great circles of $S^3$ contained in its equator, 
one deduces that a great circle on the Bloch sphere is a geodesic of $\statesof{\qubit}$, which is entirely contained in the projective space of pure states and is known as a Fubini--Study geodesic.
We call in the sequel {\textit{boundary geodesics}} the geodesics contained in the boundary of the manifold of quantum states, such as the Fubini--Study geodesics.      

If $\braket{\psi}{\phi} \not= 0$, there is up to time reversal a unique geodesic arc joining $\rv$ and $\sv$, which is a Fubini--Study geodesic~\cite{AharonovAnandan1990}. 
Actually, bulk geodesics can not pass through non-orthogonal pure states because they are given by ellipses centered at $O$ with unit major semi-axis length, intersecting the Bloch sphere at diametrically opposite points.

In contrast, when
$\ket{\psi}$ and $\ket{\phi}$ are orthogonal, 
the Bloch vectors $\rv$ and $\sv$ are diametrically opposite, thus 
there are infinitely many  great circles  on the Bloch sphere connecting  them, that is, infinitely many Fubini--Study geodesic arcs join the two orthogonal states~\cite{AharonovAnandan1990}.
Moreover, there are also infinitely many bulk geodesic arcs joining these states, having all the same length $\arccos(|\braket{\psi}{\phi}|) = \pi/2$, which is also the same as that of the Fubini--Study arcs.
In fact, there are infinitely many ellipses (\ref{eq-Bures_geodesic_qubit_states})  intersecting the Bloch sphere at $\rv$ and $\sv=-\rv$ and contained in the interior of the Bloch ball except at those points. The diameter of the Bloch ball 
connecting $\rv$ and $\sv$ is a bulk geodesic as well.
An illustration of these results is given 
 in Fig.~\ref{fig-geodesics_qubit}(c).

\vspace{2mm}

The purpose of the next sections is to generalize these results to quantum systems with higher dimensional Hilbert spaces, considering non-faithful states $\rho$ and $\sigma$ with arbitrary ranks. 
As we shall see in Secs.~\ref{sec-regularization_method} and \ref{sec-extension_method}, 
although the dichotomy between the existence of a unique or an infinite number of shortest geodesics remains valid, in higher dimensions one may find pairs of states
having non orthogonal supports and such that there are infinitely many shortest geodesic arcs connecting them.

\subsection{Bulk geodesics for general quantum systems} \label{sec_background_on_Bures_geodesics}

In this subsection we review the results of Refs.~\cite{Ericson05,Barnum_thesis,moi_geodesics_QMetrology} on the Bures geodesics connecting two faithful states. Let us first fix some notation.
Given a quantum system with a finite-dimensional Hilbert space $\Hh$, $n = \dim \Hh < \infty$, we  denote by $\states$ the set its quantum states (pure or mixed), \ie, the set of density matrices
\begin{equation}
\states   = \big\{ \rho : \Hh \to \Hh, \; ;\; \rho \geq  0 , \tr \rho =1 \big\}\;.
\end{equation}
The boundary $\partial \states$ of $\states$ consists of density matrices $\rho$ having at least one vanishing
eigenvalue, \ie, having ranks strictly smaller than $n$ (non-faithful states).
The interior of $\states$ is the open manifold $\states^\inv$ of faithful (i.e., invertible) density matrices.
The manifold $\states$ is equipped with the Bures metric.
The corresponding geodesic 
distance is the Bures arccos distance~\cite{Nielsen}
\begin{equation} \label{eq-Bures_arccos_dist}
d_\Bures (\rho ,\sigma ) = \arccos \sqrt{F(\rho,\sigma)} \;,
\end{equation}
where $\rho$ and $\sigma$ are two quantum states and
\begin{equation} \label{eq-fidelity}
F(\rho ,\sigma) =  \big( \tr | \sqrt{\sigma} \sqrt{\rho} | \big)^2\;,\;
| \sqrt{\sigma} \sqrt{\rho} | = 
( \sqrt{\rho} \, \sigma \sqrt{\rho})^\onehalf 
\end{equation}
is the fidelity. In the qubit case the fidelity is given by~(\ref{eq-fidelity_for_qubit_states}). 
As in that case, we call bulk geodesics the Bures geodesics that are contained in the interior $\states^\inv$, except possibly at a finite number of intersection points with the boundary.

If at least one of the states $\rho$ and $\sigma$ is faithful (invertible), the shortest geodesic arc joining $\rho$ and $\sigma$ is  given by~\cite{Ericson05,Barnum_thesis,moi_geodesics_QMetrology} 
\begin{eqnarray} \label{eq-Bures_geodesics_V}
\nonumber
&& 
\gamma_{\geo} ( \tau) =\frac{1}{\sin^2 \theta}
\bigg( \sin^2 ( \theta - \tau)\,  \rho  + \sin^2 ( \tau )\, \sigma
\\ 
& & 
+\sin ( \theta - \tau)  \sin ( \tau) \Big(
\sqrt{\sigma} U_{\sigma\rho} \rho^{1/2} + {\mathrm{h.c.}} \Big)  \bigg)\;,
\end{eqnarray}
$0 \leq \tau \leq \theta$, with $\theta = d_{\Bures}(\rho,\sigma)$ and $U_{\sigma \rho}$ the unitary operator in the polar decomposition
 \begin{equation} \label{eq-def_U_sigma_rho}
 \sqrt{\sigma}\sqrt{\rho} = U_{\sigma \rho} \Lambda_{\sigma \rho}
 \;,\; \Lambda_{\sigma\rho}= | \sqrt{\sigma} \sqrt{\rho} | \;.
 \end{equation}  
The geodesic $\gamma_\geo: \tau \in [0,\pi] \mapsto \gamma_\geo(\tau)$ defined by (\ref{eq-Bures_geodesics_V}) is a closed curve contained in $\states^\inv$,
save for a finite number of intersections 
with $\partial \states$. 
All geodesics bounce  $q$ times on the boundary, at states 
$\rho_i \in \partial \states$, $i=1,\ldots,q$, where $q\leq n$ is the number of distinct eigenvalues
 of the non-negative operator
 \begin{equation} \label{eq-definition_M}
 M_{\rho\sigma} = \rho^{-1/2} \Lambda_{\sigma \rho} \rho^{-1/2} 
 = \sqrt{\sigma}\, U_{\sigma \rho} \rho^{-1/2}\;.
 \end{equation}
As shown in~\cite{Ericson05,moi_geodesics_QMetrology},  the intersection states 
$\rho_i$ have kernels $P_i \Hh$,
$P_i$ being the eigenprojectors of $M_{\rho\sigma}$.
 An important consequence for what follows is that 
the states $\rho_i$ have orthogonal kernels.

Let us point out that the curve (\ref{eq-Bures_geodesics_V}) gives the shortest geodesic arc $\rho \to \sigma$. 
There are other geodesic arcs with larger lengths, bouncing at least once on the boundary between $\rho$ and $\sigma$, which have been determined in~\cite{moi_geodesics_QMetrology}. There are $2^n$ geodesic arcs from $\rho$ to $\sigma$ (counting time reversals), save in the special case where 
$\Lambda_{\sigma\rho}$ has degenerate eigenvalues, where there are infinitely many arcs, albeit the shortest one with length $\theta=d_{\Bures} (\rho,\sigma)$ is always unique. 
On the other hand, there is up to time-reversal a unique geodesic joining an invertible state $\rho$ to a state on the boundary (see Theorem 6 in~\cite{moi_geodesics_QMetrology}).
In what follows we restrict ourselves to the shortest geodesic arcs.

Let us briefly explain the method used in~\cite{Ericson05,Barnum_thesis,moi_geodesics_QMetrology} to determine the bulk geodesics (\ref{eq-Bures_geodesics_V}). It relies on a $G$-fiber-bundle approach using  purifications of the states $\rho$ as pure states on an enlarged  system-ancilla  Hilbert space $\Hh \otimes \Hh_A$:  the fiber at $\rho$ is formed by all normalized vectors $\ket{\Psi} \in \Hh \otimes \Hh_A$ such that $\rho = \tr_A \ketbra{\Psi}{\Psi}$. The fibers can be identified with the unitary group $U(n_A)$ acting on the ancilla space $\Hh_A$, where $n_A=\dim\Hh_A \geq n$. 
The geodesics $\gamma_\geo$ on the base manifold $\states$
are obtained by projecting out geodesics on the unit hypersphere of $\Hh \otimes \Hh_A$ having tangent vectors perpendicular to the fibers at each point. 

Let us point out that this approach is not easy to apply for boundary geodesics, because the boundary $\partial \states$ is stratified into sub--manifolds according to the ranks of the density matrices and the dimensions of the fibers are rank-dependent. From a mathematical viewpoint, the problem originates in the fact that 
the closed set $\states$ is not a Riemannian submersion~\cite{moi_geodesics_QMetrology}.

\subsection{Overview of the result of Sections~\ref{sec-regularization_method} and~\ref{sec-extension_method}} \label{sec_overview}

We are interested in what follows in the shortest Bures geodesic arcs  $\gamma_\geo^{1\to 2}$ joining  
two non-faithful density matrices $\rho_1 \in \partial \states$ and $\rho_2 \in \partial \states$. We write the range (support) and kernel of these matrices as
\begin{equation} \label{eq-notation_kernel_and_support}
\Hh_i  =  \supp ( \rho_i) \; ,\; \Hh_i^\bot = \ker(\rho_i) \;,\; i=1,2\;.
\end{equation}
These subspaces have dimension $r_i$ and $n-r_i$, respectively, where  $0<r_i<n$ is the rank of $\rho_i$ and $n$ is the dimension of the system Hilbert space $\Hh$.

We present in Secs.~\ref{sec-regularization_method} and~\ref{sec-extension_method} two alternative methods to determine $\gamma_\geo^{1\to 2}$. 
The first one consists in obtaining $\gamma_\geo^{1\to 2}$
from the shortest geodesic arc joining two perturbed states in the interior $\states^\inv$ of $\states$, using the results of Sec.~\ref{sec_background_on_Bures_geodesics} 
and letting the perturbed states converge to $\rho_1$ and $\rho_2$ (regularization approach). When the limiting curve is independent of the paths along which the two perturbed states approach the non-perturbed ones, we argue that there is a unique shortest geodesic arc $\gamma_\geo^{1\to 2}$. In the opposite case, each limiting curve is a possible geodesic arc $\gamma_\geo^{1\to 2}$, having  length equal to the Bures arccos  distance between $\rho_1$ and $\rho_2$ (by continuity of this distance). 
The second method consists in constructing $\gamma_\geo^{1\to 2}$ from bulk geodesics in some subspace $\Kk \subseteq \Hh$ containing $\Hh_1$ and $\Hh_2$.  
As we shall see, this method fails to give all possible geodesics arcs. For instance, Fubini--Study geodesics for pure states cannot be recovered in this way.
However, when successful the method gives well-defined arcs having support in $\Kk$, which agree with those obtained from the regularization approach. 

By combining the two  approaches and based on numerical results for qutrit and two-qubit systems, we conjecture that a necessary and sufficient condition for the existence of a unique shortest geodesic arc $\gamma_\geo^{1 \to 2}$ from $\rho_1$ to $\rho_2$ is that one or both of the following two conditions are fulfilled:
\begin{equation} \label{eq-necessary_and_sufficient_cond_unique_geodesic}
\Hh_1^\bot \cap \Hh_2=\{0\}
\quad \text{or} \quad  
\Hh_2^\bot \cap \Hh_1=\{0\}\;.	
\end{equation}
When (\ref{eq-necessary_and_sufficient_cond_unique_geodesic}) is satisfied, $\gamma_\geo^{1 \to 2}$ has rank $\max\{ r_1, r_2\}$.   
In the opposite case, there are infinitely many geodesics passing through $\rho_1$ and $\rho_2$, which are such that the  arc length from $\rho_1$ to $\rho_2$ is equal to the distance between the two states, in analogy with what happens for orthogonal pure states of a qubit. These geodesics have ranks between 
$\max\{ r_1, r_2\}$ and $\dim ( \Hh_1 + \Hh_2)$. The pure state case and the study of the qutrit and two-qubit systems (see tables~\ref{tab:geodesics_N3} and~\ref{tab:classification_geodesics}) show that geodesics $\gamma_\geo^{1\to 2}$ with different different ranks may coexist. 
Remarkably, the set of pairs of states $(\rho_1,\rho_2)$ which do not fulfill (\ref{eq-necessary_and_sufficient_cond_unique_geodesic}) is not limited to orthogonal states when $n\geq 3$ and contains pairs of states with non-orthogonal kernels when $n \geq 4$. This set is, however, of measure zero.

 \section{Geometric quantum speed limit} \label{sec_QSL}

The geometric approach to the QSL, first formulated by Anandan and Aharonov~\cite{AharonovAnandan1990}, consists in viewing time evolutions as curves in the manifold of quantum states. For a system in a pure state undergoing a unitary evolution
$\ket{\psi_t} = U_t \ket{\psi_0}$, the length of the curve 
$\gamma: t \in [0, T] \mapsto  \rho_{\psi_t}$, with $\rho_{\psi_t}=  \ketbra{\psi_t}{\psi_t}$, is given by
\begin{equation} \label{eq-FB_length_of_curve_pure_states}
\ell (\gamma) = \int_\gamma \D s_\FS = \frac{1}{\hbar} \int_0^T \sqrt{\big\langle \big(\Delta H(t) \big)^2 \big\rangle_t} \, \D t\;,
\end{equation}
where $\D s_{\FS}$ is the Fubini--Study metric  in the projective space $P \Hh$ of pure states, 
\begin{equation} \label{eq-Fubini_Study_metric}
	\D s_{\FS}^2 = \big( \| \dot{\psi_t} \|^2 - | \braket{\psi_t}{\dot{\psi}_t} |^2 \big) \D t^2 
\end{equation}
(hereafter the dot stands for the time derivative).
The second equality in (\ref{eq-FB_length_of_curve_pure_states}) involves  the variance
of the time-dependent Hamiltonian
$H(t) = \I \hbar \,\dot{U}_t U_t^\dagger$ generating the evolution, 
$\langle (\Delta H(t))^2 \rangle_t = \bra{\psi_t} H(t)^2 \ket{\psi_t} -  \bra{\psi_t} H(t) \ket{\psi_t}^2$. The equality follows from the identity $\D s_{\FS}^2 = \langle (\Delta H(t))^2 \rangle_t\, \D t^2/\hbar^2$ along the path $\gamma$, see (\ref{eq-Fubini_Study_metric}). 
Note that 
the integral in the \RHS of (\ref{eq-FB_length_of_curve_pure_states})
is a geometrical quantity, being independent of the Hamiltonian $H(t)$ used to transport the initial state $\rho_i$ into the final state $\rho_f$ along a given path $\gamma$. For instance, the transformation $H(t) \hookrightarrow H(\tau(t)) + \alpha(t) \,\identity$, where $\tau(t)$ and $\alpha(t)$ are arbitrary smooth functions such that $\dot{\tau}(t)>0$, does neither change the curve $\gamma$ nor its length $\ell (\gamma)$, so that
the integral in  (\ref{eq-FB_length_of_curve_pure_states}) remains unchanged modulo the  re-parametrization of the curve ($T \hookrightarrow  \tau (T)$, $\D t \hookrightarrow \D \tau = \tau'(t) \D t$).

The geodesic distance associated to the metric (\ref{eq-Fubini_Study_metric}) is
\begin{equation} \label{eq-Fubini-Study_distance}
	d_\FS ( \rho_{\psi}, \rho_\phi ) = \arccos | \braket{\psi}{\phi} |\;.
\end{equation}
The \RHS is independent of the choice of the representatives $\ket{\psi}$ and $\ket{\phi}$ of $\rho_\psi$ and $\rho_\phi$, \ie, of the global phase factors multiplying the normalized vectors $\ket{\psi}$ and $\ket{\phi}$.  Note that $d_\FS ( \rho_{\psi}, \rho_\phi )$ coincides with the Bures distance (\ref{eq-Bures_arccos_dist}) for pure states $\rho=\rho_\psi$ and $\sigma=\sigma_\phi$, since then 
$F(\rho_\psi,\rho_\phi)= | \braket{\psi}{\phi}|^2$ reduces to the pure state fidelity.
Since the length $\ell (\gamma)$ of the curve $\gamma$ is larger than or equal to the distance between the initial state 
$\rho_i=\gamma(0)$ and final state $\rho_f = \gamma(T)$, 
one gets the bound
\begin{equation} \label{eq-geometric_QSL_pure_states}
\frac{1}{\hbar} \int_0^T \sqrt{\langle (\Delta H(t))^2 \rangle_t}\, \D t \;\geq \; d_\FS ( \rho_i, \rho_{f})\;.
\end{equation} 
For an evolution generated by a time-independent Hamiltonian $H(t)=H$ and if the states $\rho_i$ and $\rho_f$ are orthogonal, 
$\langle (\Delta H(t))^2 \rangle_t = \Delta E^2$ is time-independent and $d_\FS ( \rho_{i}, \rho_f)=\pi/2$. Then
one recovers the TM bound  $T \geq h / 4 \Delta E$. This bound sets a limit on the minimal time $T$ for transforming the state $\rho_i$ into an orthogonal state. 
An analog bound $T \geq h / 4 \overline{\Delta E}$~\cite{AharonovAnandan1990} is obtained for a time-dependent Hamiltonian if one fixes the time-averaged energy fluctuation
\begin{equation} \label{eq-time_av_fluctuation_energy}
\overline{\Delta E} = \int_0^T \Delta E (t) \,\frac{\D t}{T}  = \int_0^T \sqrt{\langle (\Delta H(t))^2 \rangle_t} \,\frac{\D t}{T} \;.
\end{equation} 
While there exists other lower bounds on the minimal time to steer an initial state into a target state, like the Margolus--Levitin (ML) bound~\cite{MargolusLevitin1998}, 
it is important to keep in mind that the minimization of the steering time $T$ must be done under an energy constraint, reflecting the limiting resource available in a given experiment. Without such a constraint the minimal time goes to zero.
 If the constraint is on the energy fluctuation, one must fix the time-averaged energy dispersion (\ref{eq-time_av_fluctuation_energy}) and use the MT bound.
 In contrast, if the constraint is on mean energy above the ground state energy, the ML bound must be used. 
In the first case, the TM bound can be saturated. The minimal time is obtained when the curve $\gamma$ is a Fubini--Study geodesic arc. 

The preceding arguments can be generalized to mixed states in the manifold of quantum states 
$\states$ as follows~\cite{Uhlmann91,Taddei2013}. Consider a curve
$\gamma : t \in [0,T] \mapsto \rho_t$ in $\states$ joining two mixed states $\rho_i$ and $\rho_f$, where the time evolution is given by a family of Completely Positive Trace-Preserving (CPTP) maps $\Mm_t$ (quantum channels),
$\rho_t = \Mm_t (\rho_i)$. The geometric QSL bound corresponding to (\ref{eq-geometric_QSL_pure_states}) reads
\begin{equation} \label{eq-Bures_length_of_curve_mixed_states}
	\ell (\gamma)  =  \int_\gamma \D s_\Bures = \frac{1}{2} 
	\int_0^T \sqrt{\Ff_Q ( \{ \rho_t\} )}\, \D t
\geq  d_\Bures ( \rho_i, \rho_f )\;,	
\end{equation}
 where $d_\Bures$ is the arccos Bures distance (\ref{eq-Bures_arccos_dist}) and  $\Ff_Q ( \{ \rho_t\})$ is  the Quantum Fisher Information (QFI), related for faithful states to the  
Bures metric by $\D s_\Bures^2=\Ff_Q ( \{ \rho_t\})/4$.

The energy constraint behind the QSL bound  (\ref{eq-Bures_length_of_curve_mixed_states}) is more intricate than in the pure state case, as quantum evolutions of a system $S$ described by CPTP maps are not given directly in terms of some Hamiltonian. Nevertheless, one can always view these evolutions as resulting from an coupling of $S$ with an ancilla $A$, the joint system $SA$ undergoing a unitary evolution (Stinespring theorem~\cite{Nielsen}).  
Let us consider 
a curve $\Gamma: t\in [0,T] \mapsto \ket{\Psi_t}$ such that for any $t$, the normalized vector
$\ket{\Psi_t}$ is a purification of $\rho_t$ in the system-ancilla Hilbert space $\Hh \otimes \Hh_A$, that is, $\rho_t = \tr_A \ketbra{\Psi_t}{\Psi_t}$. 
Such a curve  $\Gamma$ 
is called a lift of $\gamma$. 
It is uniquely specified by a choice of  purification $\ket{\Psi_i}$ of the initial state $\rho_i$ together with a time-dependent Hamiltonian  $H_{SA} (t)$ generating the system-ancilla evolution by the Schr\"odinger equation 
$\I \hbar \ket{\dot{\Psi}_t} = H_{SA} (t) \ket{\Psi_t}$.
Hereafter, we assume that 
the ancilla Hilbert space $\Hh_A$ has dimension $n_A \geq n$. 
By an appropriate phase choice, we can restrict our analysis to lifts satisfying $\braket{\Psi_t}{\dot{\Psi}_t}=0$ for all $t$, having vanishing
energy expectation $\bra{\Psi_t} H_{SA}(t) \ket{\Psi_t}=0$.
An energy constraint associated to (\ref{eq-Bures_length_of_curve_mixed_states}) can be formulated in terms of the system-ancilla  energy dispersion $\Delta E_{SA} (t) = \hbar \| \dot{\Psi}_t\|= \big\langle H_{SA} (t)^2 \big\rangle_t^{1/2}$ minimized over all possible lifts $\Gamma$ of the evolution $\gamma$. 
More precisely, let us first note that
according to Uhlmann's theorem~\cite{Uhlmann76}, the Bures metric
  $\| \dot{\rho}_t \|_\Bures^2 = \Ff_Q ( \{ \rho_t\} )/4$ in (\ref{eq-Bures_length_of_curve_mixed_states}) can be obtained by minimizing the square norm 
 $\| \dot{\Psi}_t \|^2$ of the tangent vector $\ket{\dot{\Psi}_t}$ over all differentiable lifts of $\gamma$. Thus~\cite{Taddei2013}
\begin{eqnarray} \label{eq-geometric_QSL_for_mixed_states}
\nn
\frac{1}{\hbar} \int_0^T \sqrt{\big\langle \big(\Delta H_{SA}(t) \big)^2 \big\rangle_t} \, \D t & \geq &  
	\frac{1}{2} \int_0^T \sqrt{\Ff_Q ( \{ \rho_t\} )}\, \D t
\\
& \geq   &
d_\Bures ( \rho_i, \rho_f ) \;.
\end{eqnarray}
We now define the fiber $F_\rho \subset \Hh \otimes \Hh_A$ at $\rho \in \states$ 
as the set formed by all purifications of $\rho$. This fiber can be identified with the unitary group $U(n_A)$.
By decomposing the tangent vectors  of $\Gamma$ into their horizontal and vertical components, which are respectively orthogonal  and parallel to  $F_{\rho_t}$,  \ie, $\ket{\dot{\Psi}_t}=\ket{\dot{\Psi}_t^\hor}+\ket{\dot{\Psi}_t^\verti} $, 
  and invoking the Pythagorean theorem
  $\| \dot{\Psi}_t \|^2 = \| \dot{\Psi}_t^\hor \|^2+ \| \dot{\Psi}_t^\verti \|^2$, 
  one finds that the first inequality in (\ref{eq-geometric_QSL_for_mixed_states})
  is saturated if and only  if $\ket{\dot{\Psi}_t} = \ket{\dot{\Psi}_t^\hor}$ for all $t$~\cite{moi_geodesics_QMetrology}. A lift $\Gamma$ of $\gamma$ with this property is called a horizontal lift. 
The minimal time-average energy dispersion of the evolution 
$\gamma$ is given by
\begin{eqnarray} \label{eq-time_av_fluctuation_energy_mixed_states}
\nn
\overline{\Delta E}_\mmin 
& = & \hbar \,\min_\Gamma \int_0^T \| \dot{\Psi}_t \| \frac{\D t}{T} 
 = 
\int_0^T \sqrt{\big\langle  H_{SA}^\hor(t)^2\big\rangle_t^\hor} \,
 \frac{\D t}{T}  
\\
& = & 	\frac{\hbar}{2T} \int_0^T \sqrt{\Ff_Q ( \{ \rho_t\} )}\, \D t\;,
\end{eqnarray}
where the minimum is over all lifts $\Gamma$ of $\gamma$, $H_{SA}^\hor(t)$ is the Hamiltonian generating the horizontal lift  $t \mapsto \ket{\Psi_t^\hor}$, and $\langle O \rangle_t^\hor = \bra{\Psi_t^\hor} O \ket{\Psi_t^\hor}$. 
Then the generalized MT bound reads
\begin{equation} \label{eq-generalized_MT_bound}
T \geq \frac{\hbar\, d_\Bures ( \rho_i, \rho_f)}{\overline{\Delta E}_\mmin}\;.
\end{equation}
This bound is saturated if and only if the second inequality in (\ref{eq-geometric_QSL_for_mixed_states}) holds as an equality, which occurs when 
the evolution path $\gamma$ is a Bures geodesic arc.
For such a path the Hamiltonian $H_{SA}^\hor$ is known to be 
time-independent~\cite{moi_geodesics_QMetrology}, so that 
$\overline{\Delta E}_\mmin = \bra{\Psi_i} (H_{SA}^\hor)^2 \ket{\Psi_i}$.  

Let us stress that the QSL bound (\ref{eq-generalized_MT_bound}) gives the optimal evolution time under the constraint of a minimal time-average energy dispersion equal or lower than a fixed energy cost $\overline{\Delta E}_\mmin$. The bound derived in~\cite{Adesso2016}, which arises from monotonous Riemannian distances on $\states$ not necessarily given by the Bures distance, does not have an interpretation in terms of the aforementioned energy cost constraint. 
  
Let us consider two  pure states $\rho_i$ and $\rho_f$.
It is natural to ask the following question:
{\textit{does there exist a path $\gamma$ formed by mixed states that 
	transforms $\rho_i$ into $\rho_f$ faster, or at least as fast as, pure state evolutions, given that the time-average energy dispersions  (\ref{eq-time_av_fluctuation_energy_mixed_states}) and (\ref{eq-time_av_fluctuation_energy}) are equal\,?}  
The fact that the Bures distance 
coincides for pure states with the Fubini-Study distance 
implies, in view of (\ref{eq-geometric_QSL_pure_states}), (\ref{eq-time_av_fluctuation_energy}) and (\ref{eq-generalized_MT_bound}), that no evolution in $\states$ can be faster than a Fubini-Study geodesic. This property is intrinsically geometric. For instance, if the Bloch ball of a qubit was equipped with the Euclidean metric, instead of the Bures metric, then geodesic straight segments joining two points on the sphere, contained inside the ball, would be shorter than the corresponding geodesic arc on the sphere.
 However, geometry does not prevent mixed-state evolutions to transform $\rho_i$  into $\rho_f$ {\textit{as fast as}} optimal pure-state evolutions. Actually, in the qubit case it has been shown in Sec.~\ref{sec-geodesics_qubit} that there exists infinitely many geodesics contained in the interior of the Bloch ball
  joining two orthogonal pure states $\rho_i$ and $\rho_f$  (see Fig.~\ref{fig-geodesics_qubit}).  
 This means that optimal evolutions are not only given by  Fubini-Study geodesics, but also by rank-$2$ Bures geodesics.  
 Conversely, if $\rho_i$ and $\rho_f$ are not orthogonal, 
 \ie, $d_\Bures ( \rho_i,\rho_f) < \pi/2$, there is a unique Fubini--Study shortest geodesic arc  and thus a unique fastest evolution, and mixed-state evolution are not optimal. 
 We will prove in Sec.~\ref{sec-regularization_method} below that this result remains valid for systems with higher-dimensional Hilbert spaces. 
 
 Our results outlined in Sec.~\ref{sec_overview} show that for initial and final states $\rho_i$ and $\rho_f$ of ranks higher than one, the orthogonality of $\rho_i$ and $\rho_f$ is not necessary in order to have an infinite family of optimal paths saturating the MT bound (\ref{eq-generalized_MT_bound}).
 Remarkably, this implies that some freedom in choosing the fastest evolution to transform a non-faithful state into another  may occurs even when 
 $d_\Bures ( \rho_i,\rho_f) < \pi/2$, in contrast to what happens for pure states.
 Explicit examples for qutrit and two-qubit systems are presented in Sec.~\ref{sec-qutrit} and~\ref{sec-2qubit} below.

\section{Regularization method} \label{sec-regularization_method}

\begin{figure}[htbp]
	\begin{center}
		\includegraphics[width=7cm,angle=0]{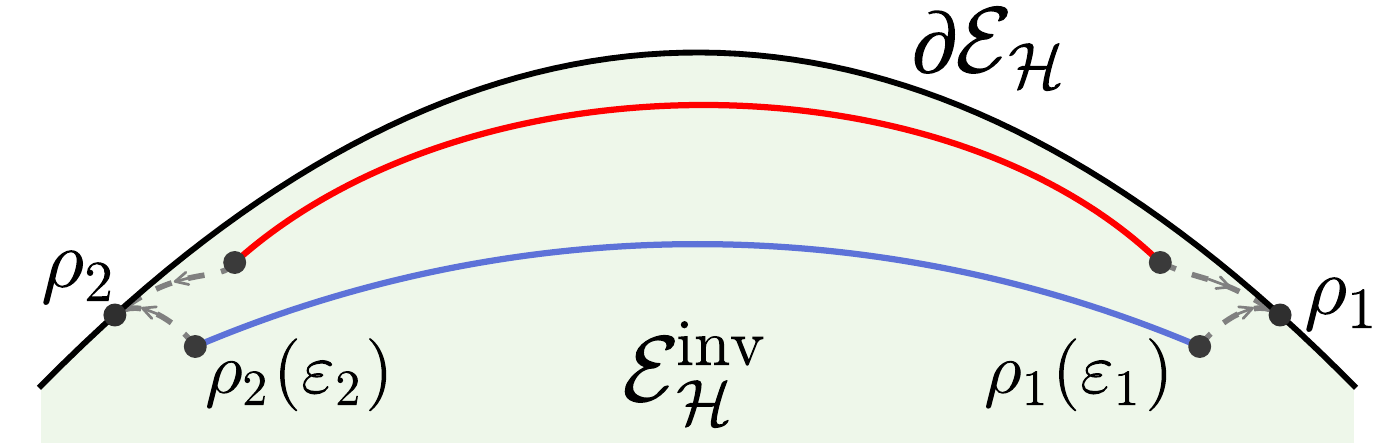}
	\end{center}  
	\captionsetup{format=plain,justification=raggedright}
	\caption{Geodesic arcs joining the perturbed states $\rho_1(\eps_1)$ and $\rho_2(\eps_2)$ of two non-faithful states $\rho_1$ and $\rho_2$  in the regularization method (plain red and blue curves). The regularization paths
	$\eps_i  \mapsto \rho_i(\eps_i)$, $i=1,2$, are shown in dashed lines.	 	
	}
	\label{fig-regularization_method}
\end{figure}

In order to determine the geodesics $\gamma_\geo^{1\to 2}$ we rely on the following assumption.

\vspace{2mm}

\noindent {\bf Hypothesis 1:} {\it 
	All the shortest geodesics  joining two non-invertible states $\rho_1$ and $\rho_2$ can be obtained as limits when $\eps_1,\eps_2 \to 0+$ of the shortest geodesic $\gamma_{\geo}^{1 \to 2}(\eps_1, \eps_2)$ joining the  perturbed invertible states $\rho_1(\eps_1)$ and $\rho_2(\eps_2)$, where $\eps_i \in  (0,1] \mapsto \rho_i(\eps_i)$ are smooth curves in $\states^\inv$ with limits $\rho_i(0+)=\rho_i$, $i=1,2$.   
}

\vspace{2mm}

The shortest geodesic  $\gamma_{\geo}^{1 \to 2}(\eps_1, \eps_2)$, being contained in 
$\states^\inv$, is unique and given by (\ref{eq-Bures_geodesics_V}). In order to determine its limit, it is enough to 
obtain the limit of the unitary $U_{\rho_2(\eps_2) \rho_1(\eps_1)}$ in the polar decomposition of $\sqrt{\regrhotwo}\sqrt{\regrhoone}$ (see~(\ref{eq-def_U_sigma_rho})),
\begin{equation} \label{eq-regularized_unitary_21_def}
U_{\rho_2\rho_1}^{\reg} = \lim_{\eps_1,\eps_2 \to 0} U_{\rho_2(\eps_2) \rho_1(\eps_1)}\;.
\end{equation}
We shall prove below that the limit exists. This limit is regularization-dependent, \ie, it depends on the choice of the curves  $\eps_i \mapsto \rho_i(\eps_i)$.
Replacing $\rho$ by $\rho_1(\eps_1)$ and $\sigma$ by $\rho_2(\eps_2)$ in (\ref{eq-Bures_geodesics_V}), taking the limit $\eps_1,\eps_2 \to 0$ and using 
$\rho_i(\eps_i) \to \rho_i$ and $d_{\Bures} ( \rho_1(\eps_1),\rho_2(\eps_2)) \to \theta = d_{\Bures} ( \rho_1,\rho_2)$, one obtains
\begin{eqnarray} \label{eq-Bures_geodesics_reg}
\nonumber
&& 
\gamma_{\geo}^{1 \to 2} ( \tau) =\frac{1}{\sin^2 \theta}
\bigg( \sin^2 ( \theta - \tau)\,  \rho_1  + \sin^2 ( \tau )\, \rho_2
\\ 
& & 
+\sin ( \theta - \tau)  \sin ( \tau) \Big(
\sqrt{\rho_2}\, U_{\rho_2\rho_1}^{\reg} \sqrt{\rho_1} + {\mathrm{h.c.}} \Big)  \bigg)\;.
\end{eqnarray}

We now determine $U_{\rho_2\rho_1}^\reg$. To simplify notation we write 
$U_{2 1}(\veceps)= U_{\regrhotwo \,\regrhoone}$ and 
$\Lambda_{2 1}(\veceps)=| \sqrt{\regrhotwo}\,\sqrt{\regrhoone}|$,
with $\veceps= ( \eps_1,\eps_2)$. 
Let us introduce the spectral decompositions
\begin{eqnarray*} \label{eq-spectral_dec_Lambda}
\nn
\Lambda_{2 1}^2(\veceps) \! =\! \sqrt{\regrhoone} \regrhotwo \sqrt{\regrhoone}
& \!\!=\!\! & \sum_{k=1}^n \lambda_k^2(\veceps ) 
\ketbra{ u_k (\veceps)}{u_k(\veceps)},
\\
\Lambda_{1 2}^2 (\veceps) = \sqrt{\regrhotwo} \regrhoone \sqrt{\regrhotwo} 
& \!\! = \!\! & 
\sum_{k=1}^n \lambda_k^2(\veceps ) 
\ketbra{ v_k (\veceps)}{v_k(\veceps)}\;.
\end{eqnarray*}
Recall that the polar decomposition $A= U |A|$ implies $A A^\dagger = U |A|^2 U^\dagger$. Hence
\begin{equation} \label{eq-polar_dec_bis}
\Lambda_{12}^2 (\veceps ) = U_{21} ( \veceps) \Lambda_{21}^2 (\veceps) U_{21}^\dagger (\veceps)\;.
\end{equation}
Thus  $U_{21} (\veceps)$ transforms the eigenvectors $\ket{u_k(\veceps)}$ of $\Lambda_{21}^2(\veceps)$ into eigenvectors $\ket{v_k(\veceps)}$ of $\Lambda_{12}^2(\veceps)$ with the same eigenvalue $\lambda_k(\veceps)$, \ie,
\begin{equation}
U_{21} (\veceps) = \sum_{k=1}^n \ketbra{v_k(\veceps)}{u_k ( \veceps)}\;.
\end{equation}
One can determine the eigenvectors $\ket{u_k( \veceps)}$ and $\ket{v_k (\veceps )}$ by using perturbation theory. Since we are interested in the limit $\eps_1, \eps_2 \to 0$, one only needs these eigenvectors to zeroth order, 
$\ket{u_k( \veceps)}= \ket{u_k^{(0)}(\veceps)}+ O (|\veceps|)$ 
	and  $\ket{v_k (\veceps )}= \ket{v_k^{(0)}} + O (|\veceps|)$. Then (\ref{eq-regularized_unitary_21_def}) reads
\begin{equation} \label{eq-regularized_unitary_21}
U_{\rho_2\rho_1}^\reg = \lim_{\eps_1,\eps_2 \to 0} U_{21} ( \veceps)
= \sum_{k=1}^n \ketbra{v_k^{(0)} }{u_k^{(0)}}\;.
\end{equation}
Let us introduce the projectors
\begin{equation}
\begin{array}{ll}
\Pi_i  & = \text{ orthogonal projector onto }\Hh_i=\supp ( \rho_i)\;,
\\
\Pi_{\rho_i\rho_j}  &  = \text{ orthogonal projector onto } 
\supp \Lambda_{\rho_i\rho_j}\;, \\
\Pi_{\rho_i\rho_j}^\bot & = \text{ orthogonal projector onto } 
\ker \Lambda_{\rho_i\rho_j}\;.
\end{array}
\end{equation}
with $i,j=1,2$, $i \not=j$. Hereafter, when adding a  subscript $\bot$ to a projector we refer to the projector onto the orthogonal subspace, e.g. $\Pi_{\rho_i\rho_j}^\bot = \identity - \Pi_{\rho_i\rho_j}$.

We first assume that $\Lambda_{\rho_2\rho_1}$ and $\Lambda_{\rho_1\rho_2}$ have non-degenerate eigenvalues save for the zero eigenvalue. Let us order the non-zero eigenvalues as $\lambda_1> \ldots > \lambda_m > 0$, where $m$ is the rank of 
$\Lambda_{\rho_2\rho_1}$, or equivalently, of $\Lambda_{\rho_1\rho_2}$. Standard
perturbation theory yields $\ket{u_k^{(0)}} = \ket{u_k}$ and 
$\ket{v_k^{(0)}} = \ket{v_k}$ for any $k=1,\ldots,m$.	
On the other hand, since the zero eigenvalue of 
$\Lambda_{\rho_1\rho_2}$ and $\Lambda_{\rho_2\rho_1}$
is degenerate, the zeroth order eigenvectors $\ket{u_k^{(0)}}$ and 
$\ket{v_k^{(0)}}$ for $k=m+1,\ldots,n$  must be obtained by diagonalizing
\begin{eqnarray} \label{eq-perturbations_of_Lambda}
\nn
\delta \Lambda_{21}(\veceps) & = & \Pi_{\rho_2\rho_1}^\bot ( \Lambda_{21}(\veceps)- \Lambda_{\rho_2\rho_1} ) \Pi_{\rho_2 \rho_1}^\bot \\
\delta \Lambda_{12}(\veceps) & = & \Pi_{\rho_1\rho_2}^\bot ( \Lambda_{12}(\veceps)- \Lambda_{\rho_1\rho_2} ) \Pi_{\rho_1\rho_2}^\bot\;,
\end{eqnarray} 
respectively. Because the perturbation $\delta \Lambda_{ij} ( \eps)$ depends on the two curves
$\eps_i \mapsto \rho_i(\eps_i)$, $\ket{u_k^{(0)}}$ and 
$\ket{v_k^{(0)}}$ depend on the regularization for $m<k\leq n$, whereas they are regularization-independent for $1 \leq k \leq m$.

Since $\{ \ket{u_k^{(0)}} \}_{k=1}^n$ and $\{ \ket{v_k^{(0)}} \}_{k=1}^n$ are orthonormal bases, one deduces from (\ref{eq-regularized_unitary_21}) that
$U_{\rho_2\rho_1}^\reg$ has the form
\begin{eqnarray} \label{eq-decomp_U_21^reg}
\nn
U_{\rho_2\rho_1}^\reg & = & \Pi_{\rho_1\rho_2} U_{\rho_2 \rho_1}^\reg \Pi_{\rho_2\rho_1}
+ \Pi_{\rho_1\rho_2}^\bot U_{\rho_2 \rho_1}^\reg \Pi_{\rho_2\rho_1}^\bot
\\
& = & U_{\rho_2 \rho_1}^{(0)} + U_{\rho_2 \rho_1}^{\perp}\;,
\end{eqnarray}
where
\begin{equation} \label{eq-definition_U_21^(0)}
U_{\rho_2 \rho_1}^{(0)} = \Pi_{\rho_1\rho_2} U_{\rho_2 \rho_1}^\reg \Pi_{\rho_2\rho_1} = 
\sum_{k=1}^m \ketbra{v_k}{u_k}
\end{equation}
only depends on the eigenvectors of the unperturbed operators $\Lambda_{\rho_2\rho_1}$ and $\Lambda_{\rho_1\rho_2}$. 
Recall that while the unitary $U$ in the polar decomposition $A=U|A|$ of  a non-invertible operator $A$ is not uniquely defined, its restriction 
$\supp (|A|) \to \supp ( |A^\dagger|)$ is uniquely determined.  
Hence $U_{\rho_2 \rho_1}^{(0)}$ is regularization-independent and uniquely determined from the polar decomposition
\begin{equation} \label{eq-polar_decomp_non_invertible_states}
\sqrt{\rho_2}\sqrt{\rho_1} = U_{\rho_2\rho_1}^{(0)} \Lambda_{\rho_2\rho_1}\;. 
\end{equation}
In contrast, the second contribution in (\ref{eq-decomp_U_21^reg}),
\begin{equation}
U_{\rho_2\rho_1}^\perp = \sum_{k>m}^n \ketbra{v_k^{(0)}}{u_k^{(0)}}\;,
\end{equation}
is regularization-dependent.
It is shown in Appendix~\ref{sec-Lambda_21_degenerate_eigenvalues} that these conclusions remain valid when 
$\Lambda_{\rho_2\rho_1}$ has degenerate positive eigenvalues.

We now determine the kernels and supports of $\Lambda_{\rho_2\rho_1}$ and $\Lambda_{\rho_1\rho_2}$. The former are given by
\begin{eqnarray} \label{eq-kernel_Lambda21}
\nn
\ker (\Lambda_{\rho_2 \rho_1}) 
& = & \ker ( \sqrt{\rho_2} \sqrt{\rho_1} ) = \Hh_1^\bot \oplus S_{21}\;,
\\
\ker (\Lambda_{\rho_1 \rho_2}) 
& = & \ker ( \sqrt{\rho_1} \sqrt{\rho_2} ) = \Hh_2^\bot \oplus S_{12}\;,
\end{eqnarray}
where $\Hh_i^\bot = \ker( \rho_i)$ and %
\begin{eqnarray} \label{eq-definition_S_12}
\nn
S_{21} & = & \rho_1^{-1/2} (\Hh_2^\bot \cap \Hh_1 )\;\subseteq \; \Hh_1\; ,
\\
S_{12} & = & \rho_2^{-1/2} (\Hh_1^\bot\cap\Hh_2 ) \; \subseteq \; \Hh_2
\;.
\end{eqnarray}
Thanks to the self-adjointness of $\Lambda_{\rho_2\rho_1}$ and the relation $(E\oplus F)^\bot = E^\bot \cap \,F^\bot$ for any two subspaces
 $E,F \subseteq \Hh$, the supports of $\Lambda_{\rho_2\rho_1}$ and $\Lambda_{\rho_1\rho_2}$ are given by 
\begin{eqnarray} \label{eq-supports_Lambda}
\nn
\supp ( \Lambda_{\rho_2\rho_1})  &  = & 
\Hh_1 \cap \, S_{21}^\bot = \sqrt{\rho_1}\,\Hh_2 \;\subseteq \; \Hh_1\; ,
\\
\supp ( \Lambda_{\rho_1\rho_2})  & = &  
\Hh_2 \cap \, S_{12}^\bot = \sqrt{\rho_2}\,\Hh_1 \;\subseteq \; \Hh_2\;.
\end{eqnarray}
Note that 
$m=\rank ( \Lambda_{\rho_2\rho_1})= \rank (\Lambda_{\rho_1\rho_2}) = r_2 - s_{12} = r_1 - s_{21}$, where  $s_{ij} = \dim(S_{ij})$ (the equality of the ranks follows from (\ref{eq-polar_dec_bis}) with $\veceps=0$).
This yields the  relation
\begin{equation} \label{eq-relation_dimensions_S}
r_1 + s_{12} = r_2 + s_{21}\;.
\end{equation}
Furthermore, one has in view of (\ref{eq-kernel_Lambda21})
\begin{equation} \label{eq-decomp_Pi_i}
\Pi_1=\Pi_{\rho_2 \rho_1}+\Pi_{S_{21}}\;,\;
\Pi_2=\Pi_{\rho_1\rho_2}+\Pi_{S_{12}}\;,
\end{equation}
where $\Pi_{S_{ij}}$ is the orthogonal projector onto $S_{ij}$.

Using (\ref{eq-decomp_Pi_i}), (\ref{eq-decomp_U_21^reg}) and the fact that  $S_{ij} \bot \supp ( \Lambda_{\rho_i\rho_j})$, see (\ref{eq-supports_Lambda}), 
one may write
\begin{align} \label{eq-sandwiched_unitary_U_21}
\nn
& \sqrt{\rho_2}\,U^{\mathrm{reg}}_{\rho_2\rho_1}\sqrt{\rho_1}
=
\sqrt{\rho_2}\,\Pi_2 U^{\mathrm{reg}}_{\rho_2\rho_1}\Pi_1 \sqrt{\rho_1}
\\ \nn
& \quad = \sqrt{\rho_2}\,(\Pi_{\rho_1\rho_2}+\Pi_{S_{12}})
U^{\mathrm{reg}}_{\rho_2\rho_1}
(\Pi_{\rho_2 \rho_1}+\Pi_{S_{21}})\,\sqrt{\rho_1}
\\
& \quad = \sqrt{\rho_2}\, U^{(0)}_{\rho_2\rho_1} \sqrt{\rho_1}
+ \sqrt{\rho_2}\,\Pi_{S_{12}} U^{\perp}_{\rho_2\rho_1} \Pi_{S_{21}} \sqrt{\rho_1}\;.
\end{align}
Replacing this formula into (\ref{eq-Bures_geodesics_reg}) yields
\begin{eqnarray} \label{eq-Bures_geodesics_reg_bis}
	\nonumber
	&& 
	\gamma_{\geo}^{1 \to 2} ( \tau) =\frac{1}{\sin^2 \theta}
	\bigg( \sin^2 ( \theta - \tau)\,  \rho_1  + \sin^2 ( \tau )\, \rho_2
	\\ \nonumber
	& & 
	+\sin ( \theta - \tau)  \sin ( \tau) \Big(
	\sqrt{\rho_2} \, U_{\rho_2\rho_1}^{(0)} \sqrt{\rho_1} + 
	\sqrt{\rho_2} \, R_{\rho_2\rho_1} \sqrt{\rho_1} 
	\\
	& & 
	+ {\mathrm{h.c.}} \Big)  \bigg)
\end{eqnarray}
with  $\theta = d_{\Bures} ( \rho_1,\rho_2)$ and
\begin{equation} \label{eq-def_R_rho2_rho1}
R_{\rho_2\rho_1} = \Pi_{S_{12}} U^{\reg}_{\rho_2\rho_1} \Pi_{S_{21}}
= \lim_{\eps_1,\eps_2 \to 0} 
\Pi_{S_{12}} U_{\rho_2(\eps_2)\rho_1(\eps_1)}  \Pi_{S_{21}}\;.
\end{equation}
Note that  $R_{\rho_2\rho_1}^\dagger R_{\rho_2\rho_1} \leq \Pi_{S_{21}}$ and 
$R_{\rho_2\rho_1} R_{\rho_2\rho_1}^\dagger \leq \Pi_{S_{12}}$, so that 	$R_{\rho_2\rho_1}$ is a contraction $S_{21} \to S_{12}$.  We obtain then the following theorem: 

\vspace{2mm}

\noindent {\bf Theorem 1.}
{\it A sufficient condition to have a unique regularized geodesic joining $\rho_1$ and $\rho_2$, \ie\, for the limit $\gamma_\geo^{1 \to 2} (\tau) = \lim_{\eps_1,\eps_2 \to 0+}
\gamma_{\geo}^{1 \to 2}(\eps_1,\eps_2)$ to be independent of the 
 regularization curves $\eps_ \mapsto \rho_i(\epsilon_i) $, is
\begin{equation}\label{eq-sufcond0}
S_{12}=\{0\}
\quad\text{or}\quad
S_{21}=\{0\},
\end{equation}
which is equivalent to
\begin{equation}\label{eq-sufcond1}
\Hh_1^\bot \cap\Hh_2 =\{0\}
\quad\text{or}\quad
\Hh_2^\bot  \cap\Hh_1=\{0\}\;.
\end{equation}
In that case, the geodesic is given by (\ref{eq-Bures_geodesics_reg_bis})
with $R_{\rho_2\rho_1}=0$ and $U_{\rho_2\rho_1}^{(0)}: \supp (\Lambda_{\rho_2 \rho_1}) \to \supp ( \Lambda_{\rho_1 \rho_2})$ the uniquely defined partial isometry in the polar decomposition (\ref{eq-polar_decomp_non_invertible_states}).
}

\vspace{2mm}

Note that due to (\ref{eq-relation_dimensions_S}), one of the two subspaces $S_{ij}$ is always non trivial  if $r_1\not=r_2$ and one has
\begin{enumerate}
	\item \label{case-1} if $r_1>r_2$ then (\ref{eq-sufcond0}) $\Leftrightarrow \,S_{12} = \{0\}$;
	\item \label{case-2} if $r_1=r_2$ then (\ref{eq-sufcond0}) $\Leftrightarrow S_{12} = \{0\} \Leftrightarrow S_{21} = \{0\}$;
	\item  \label{case-3} if $r_1<r_2$ then (\ref{eq-sufcond0})
	$\Leftrightarrow \,S_{21} = \{0\}$.
\end{enumerate}
Let us point out that condition (\ref{eq-sufcond0}) is generically satisfied, namely, 
the set of pairs of states $(\rho_1,\rho_2)$ which do not fulfill the 
condition is of measure zero for the Haar measure. 
Indeed, in case~\ref{case-1}, the co-dimension of $\Hh_1^\bot$ being strictly larger than the dimension of $\Hh_2$, the intersection of the two subspaces (and thus $S_{12}$) is, generically, reduced to $\{0\}$. The same is true in case~\ref{case-3} by interchanging $\rho_1$ and $\rho_2$. In case~\ref{case-2}, ${\rm{codim}} (\Hh_1^\bot)=\dim (\Hh_2)$, so that again $S_{12}=\{0\}$ holds generically.     

\vspace{2mm}

It is not easy to derive explicit expressions of the operator $R_{\rho_2\rho_1}$ for specific regularization curves $\eps_i \mapsto \rho_i (\eps_i)$. 
For instance, using $\rho_i ( \eps_i) = (1-\eps_i^2) \rho_i + \eps_i^2 \Pi_i^\bot /(n-r_i)$, the  determination of the zeroth order eigenvectors $\ket{u_k^{(0)}}$ and $\ket{v_k^{(0)}}$ of  $\Lambda_{21}^2(\veceps)$ and
$\Lambda_{12}^2(\veceps)$ with vanishing eigenvalues in the limit $\eps \to (0,0)$ requires expanding the perturbations (\ref{eq-perturbations_of_Lambda}) up to order four in $\eps_1$ and $\eps_2$ and using fourth-order perturbation theory (in fact, the first, second and third order contributions vanish).
As a consequence, we are not able to show that any contraction $R_{\rho_2\rho_1}$ from 
$S_{21}$ to $S_{12}$ can be obtained
by choosing the regularization curves $\eps_i \mapsto \rho_i ( \eps_i)$ appropriately.  
We study numerically in Appendix~\ref{sec-regularization_geodesics_pure_states}
the limit (\ref{eq-def_R_rho2_rho1}) for pure states $\rho_1$ and $\rho_2$ and find numerical evidence  that
for any contraction $R_{21} : S_{21} \to S_{12}$, one can find some regularization curves $\epsilon_i \mapsto \rho_i(\eps_i)$ such that  
$R_{21}=R_{\rho_2\rho_1}$. This implies that  the family of geodesics joining $\rho_1$ and $\rho_2$ is given by 
(\ref{eq-Bures_geodesics_reg_bis}) with
 $R_{\rho_2\rho_1}$ an arbitrary contraction $S_{21} \to S_{12}$.
Numerical investigation for states $\rho_1$, $\rho_2$ with higher ranks also indicate that  when (\ref{eq-sufcond1}) is not satisfied,  the operators $R_{\rho_2\rho_1}$  are regularization-dependent, so that the limit $\lim_{\eps_1,\eps_2 \to 0+} \gamma_{\geo}^{1 \to 2}(\eps_1,\eps_2)$ depends on the choice of the regularization curves.
Based on Theorem 1 and on these results, we make the following conjecture:

\vspace{1mm}

\noindent {\bf{Conjecture:}}
{\it Condition (\ref{eq-sufcond1}) is necessary and sufficient for the existence of a unique geodesic $\gamma_\geo^{1 \to 2}$ joining two non-invertible states $\rho_1$ and $\rho_2$. If this condition is not satisfied there are infinitely many geodesics $\gamma_\geo^{1 \to 2}$.}

\vspace{2mm}

\noindent {\bf{Pure state case:}} 

In the case where $\rho_1 = \ketbra{\psi_1}{\psi_1}$ and   $\rho_2 = \ketbra{\psi_2}{\psi_2}$ 
are pure states, one has $\Hh_j = \complex \ket{\psi_j}$ and  
 $\Hh_{i}^\bot \cap \Hh_j \not= \{0\}  \Leftrightarrow \braket{\psi_i}{\psi_j}=0$, $i \not= j$.
Thus  if $\ket{\psi_1}$ and $\ket{\psi_2}$ are not orthogonal, then from Theorem~1 there  is a unique 
geodesic joining $\rho_1$ and $\rho_2$. This geodesic is a Fubini--Study geodesic on the projective space $P \complex^n$ of pure states. Indeed, (\ref{eq-polar_decomp_non_invertible_states}) reads 
\begin{equation}
\sqrt{\rho_2}\sqrt{\rho_1}  = \ketbra{\psi_2}{\psi_1} \braket{\psi_2}{\psi_1} = U_{\rho_2\rho_1}^{(0)} \ketbra{\psi_1}{\psi_1} \, | \braket{\psi_1}{\psi_2} |\;,
\end{equation}
\ie
\begin{equation} \label{eq-Berry_parralel_transport}
U_{\rho_2\rho_1}^{(0)} \ket{\psi_1}= \E^{-\I \arg\braket{\psi_1}{\psi_2}} \ket{\psi_2}\;.
\end{equation}
This formula is nothing but the parallel transport formula in $P \complex^n$.
Substituting (\ref{eq-Berry_parralel_transport}) into  (\ref{eq-Bures_geodesics_reg_bis}) one finds that 
$\gamma^{1 \to  2}_\geo ( \tau)= \ketbra{\psi_\geo (\tau)}{\psi_\geo (\tau)}$ with
\begin{equation} \label{eq-Fubini_Study_geodesic}
\ket{\psi_\geo (\tau)}=\frac{\sin(\theta-\tau)}{\sin\theta} \ket{\psi_1}+\frac{\sin \tau}{\sin \theta} \E^{-\I \arg \braket{\psi_1}{\psi_2}} \ket{\psi_2}\;,
\end{equation}
$0 \leq \tau \leq \theta$, where $\theta = \arccos |\braket{\psi_1}{\psi_2}|$. 
Eq. (\ref{eq-Fubini_Study_geodesic}) coincides with the Fubini-Study geodesic.

On the other hand, when $\braket{\psi_1}{\psi_2}=0$, then $\Lambda_{\rho_2\rho_1}=0$ and $S_{ij} = \complex \ket{\psi_j}$, so that the regularized unitary 
(\ref{eq-decomp_U_21^reg}) reduces to its second term. Eq. (\ref{eq-sandwiched_unitary_U_21}) yields $\sqrt{\rho_2 }\,U_{\rho_2\rho_1}^\reg \sqrt{\rho_1} = z_{21} \ketbra{\psi_2}{\psi_1}$ with $z_{21} = \bra{\psi_2} U_{\rho_2\rho_1}^\bot \ket{\psi_1}$. Thus, as shown in Appendix~\ref{sec-regularization_geodesics_pure_states},
there are infinitely many geodesics given by 
\begin{align} \label{eq-geodesics_joining_pure_states_regularization}
\nn
& \gamma^{1 \to  2}_\geo ( \tau)=  \cos^2 ( \tau) \ketbra{\psi_1}{\psi_1} + \sin^2 ( \tau ) \ketbra{\psi_2}{\psi_2}  \\
& \quad + \cos(\tau)\sin(\tau) \big( z_{21} \ketbra{\psi_2}{\psi_1} +  {z}_{21}^\ast \ketbra{\psi_1}{\psi_2}\big)\;,
\end{align} 
$z_{21}$ being an arbitrary complex number of modulus smaller than one.
All these geodesics are supported in $\Span \{ \ket{\psi_1}, \ket{\psi_2}\}$; when restricted to this subspace, they may be considered as geodesics of a qubit, $\ket{\psi_1}$ and $\ket{\psi_2}$ being identified with the north and south pole of the Bloch sphere.   
As illustrated in Fig.~\ref{fig-geodesics_qubit},
for $|z_{21}|=1$ one obtains an infinity of Fubini--Study geodesics formed by the arcs of great circles joining the two poles. For $0<|z_{21}|<1$ one has 
an infinity of arcs of ellipses joining the two poles, contained in the interior of the ball. For $z_{12}=0$ one obtains the  straight vertical diameter, corresponding to a geodesic formed by commuting mixed states.

\section{Subspace restriction method} \label{sec-extension_method}

In this section we present an alternative method to determine the geodesics connecting 
non-faithful states. This method relies on the characterization of the family of geodesics passing through two states having orthogonal kernels in the subspace spanned by these two states.
We introduce  the notions of support and rank to classify these geodesics.

Given a subspace $\Kk \subset \Hh$ of dimension $k<n$ containing the supports of $\rho_1$ and $\rho_2$, consider a smooth curve  $\gamma_{\Kk}: \tau \in [0,\theta] \mapsto \gamma_{\kappa} (\tau) \in \statesof{\Kk}$ in the manifold $\statesof{\Kk}$ of quantum states  with Hilbert space $\Kk$ joining  $\rho_1 |_\Kk$ and $\rho_2 |_\Kk$. 
Then  $\gamma_{\Kk}$ can be identified with  
a curve $\widehat{\gamma}_{\Kk}$ in $\partial \states$, obtained by extending the density operators $\gamma_{\Kk}(\tau): \Kk \to \Kk$ as operators $\widehat{\gamma}_{\Kk}: \Hh \to \Hh$ with  supports contained in $\Kk$. 

In the sequel, we say that a curve 
$\gamma : \tau \in [0,\theta] \mapsto \gamma (\tau)$ in $\states$ has support $\Kk$
when the span of the union of the supports of $\gamma (\tau)$ for all  $\tau \in [0,\theta]$ is equal to $\Kk$.
We say that $\gamma$ has rank $k$ when 
$\rank (\gamma (\tau))= k$ for all $\tau\in [0,\theta]$ save for a finite number of times $0\leq t_1<\ldots< t_q \leq \theta$. 
Clearly, if $k<n$ then such a curve is contained in the boundary $\partial \states$ of $\states$. 
For an analytic curve $\gamma$, $\rank (\gamma (\tau))$ is constant as one moves along the curve excepted possibly at some discrete times $0 \leq \tau_1 < \ldots < \tau_q \leq \theta$, where one or more positive eigenvalues of $\gamma ( \tau)$ vanish and $\rank ( \gamma(\tau))$ reaches a lower value. (Actually, if e.g. $\gamma (\tau)$ has an eigenvalue $p_k(\tau)$ such that $p_k(\tau)>0$ for $0 < \tau<\tau_1$ and $p_k(\tau_1)=0$ then, by non-negativity and analyticity of the eigenvalue, $p_k$ has a strict local minimum at $\tau_1$; thus $\rank (\gamma (\tau))$ is constant in $I_1 \setminus \{ \tau_1\}$, $I_1$ being a small interval centered on $\tau_1$.) Hence the rank of an analytic curve $\gamma$ is equal or larger than the rank of any of its points $\gamma(t)$.
For instance, the geodesics (\ref{eq-Bures_geodesics_V}) joining an invertible state $\rho \in \states^\inv$ to an (invertible or non-invertible) state $\sigma \in \states$ are rank-$n$ geodesics.
According to the terminology of Sec.~\ref{sec-bulk_and_boundary_geodesics}, bulk geodesics are rank-$n$ geodesics. 

Our second method relies on the following assumption:

\vspace{2mm}

\noindent {\bf{Hypothesis 2:}} {\it If $\gamma_{\geo , \Kk}$ is a geodesic in 
$\statesof{\Kk}$ then its extension $\widehat{\gamma}_{\geo, \Kk}$ is a geodesic 
in $\partial \states$ with support in $\Kk$.}

\vspace{2mm}

According to the results of~\cite{Ericson05,moi_geodesics_QMetrology} mentioned after (\ref{eq-definition_M}), in order that there exists a rank-$n$ geodesic joining two states $\rho_1\in \partial \states$ and $\rho_2\in \partial \states$, one must have
\begin{equation} \label{eq-orthogonality_condition}
\Hh_1^\bot \; \perp \; \Hh_2^\bot \;.
\end{equation} 
Let us introduce the subspace
\begin{equation} \label{eq-def_H_12}
\Hh_{12}  =  \Hh_1 + \Hh_2\;,
\end{equation}
which has dimension
 $n_{12} = r_1+r_2 - \dim( \Hh_1 \cap \Hh_2)$. 
 Let us assume that $n_{12} <n$. Then for any subspace  $\Kk \supsetneq \Hh_{12}$ containing and not equal to $\Hh_{12}$, the subspaces $\Hh_1^\bot \cap \Kk$ and $\Hh_2^\bot \cap \Kk$ are not orthogonal, since 
 $( \Hh_1^\bot \cap \Hh_2^\bot ) \cap \Kk = \Hh_{12}^\bot \cap \Kk$ 
 is non-trivial. 
 In view of Hypothesis 2, applying
 condition (\ref{eq-orthogonality_condition}) to the states
  $\rho_i |_\Kk \in \partial \statesof{\Kk}$ restricted to the $\Kk$-subspace, one concludes that there are no geodesics 
 with support $\Kk$ and rank $k> n_{12}$ joining $\rho_1$ and $\rho_2$.  
 
 One concludes that:
 
 \vspace{2mm}
 
  {\it{All geodesics $\gamma_\geo^{1\to 2}$ joining $\rho_1$ and $\rho_2$ have supports contained in $\Hh_{12}$. In particular, their ranks satisfy}}
  \begin{equation} \label{eq-inequlities_rank_geodesics}
  	\max \{ r_1, r_2 \} \leq \rank ( \gamma_\geo^{1\to 2}) \leq n_{12}\,.
\end{equation}
 This means that when studying geodesics passing through $\rho_1$ and $\rho_2$, one can without loss of generality restrict the system Hilbert space to the subspace $\Hh_{12}$ given by (\ref{eq-def_H_12}).
Note that the same conclusion can be inferred from the regularization approach, see (\ref{eq-Bures_geodesics_reg}).

This leads us to consider the following condition, which is more general than 
(\ref{eq-orthogonality_condition}),
\begin{equation} \label{eq-restricted_orthogonality_cond}
\Hh_1^\bot \cap \Hh_{12}\;\perp\; \Hh_2^\bot \cap \Hh_{12}\;.
\end{equation} 
We distinguish the following cases: 
\begin{itemize}
	\item[(i)] If (\ref{eq-restricted_orthogonality_cond}) does not hold then  there are no rank-$n_{12}$ geodesics, that is,  all geodesics $\gamma^{1\to 2}_\geo$ have  supports strictly contained in $\Hh_{12}$.
	In fact, there are no geodesic arcs in $\statesof{\Hh_{12}}$ connecting $\rho_1|_{\Hh_{12}}\in \partial \statesof{\Hh_{12}}$ and $\rho_2|_{\Hh_{12}} \in \partial \statesof{\Hh_{12}}$.
	Then one can not determine the geodesics from Theorem~2 below.   
	
	\item[(ii)] 
If $\Hh_2 \subseteq \Hh_1$ there is a unique geodesic $\gamma_\geo^{1 \to 2}$, which has support $\Hh_1$ and rank $r_1$. In fact, in such a case $\Hh_{12}= \Hh_1$ and $n_{12}=r_1$.  
Since $\rho_1 |_{\Hh_{1}}$ is an invertible state, there is a unique rank-$r_1$ shortest geodesic arc joining $\rho_1|_{\Hh_1}$ and $\rho_2|_{\Hh_2}$ in $\statesof{\Hh_1}$, see Sec.~\ref{sec_background_on_Bures_geodesics}. By Hypothesis~2, its extension is a geodesic $\rho_1 \to \rho_2$ with support $\Hh_1$.
	On the other hand, by (\ref{eq-inequlities_rank_geodesics}) 
	there are no geodesics $\gamma_\geo^{1 \to 2}$ with rank smaller or larger than $r_1$, \ie, with support $\Kk \neq \Hh_1$.
	Similarly, when $\Hh_1 \subseteq \Hh_2$, there is a unique  
	$\gamma_\geo^{1 \to 2}$, which has rank $r_2$ and support $\Hh_2$.
		
	\item[(iii)] If (\ref{eq-restricted_orthogonality_cond}) holds and 
\begin{equation} \label{eq-non_inculsion_condition}
\Hh_2 \not\subseteq \Hh_1\;\; , \;\; \Hh_1 \not\subseteq \Hh_2\;,
\end{equation}
 we rely on the following result.
	Note that  (\ref{eq-orthogonality_condition}) implies $r_1+r_2\geq n$.  	
	
\end{itemize}

\noindent {\bf Theorem 2.}  {\it Assume that the orthogonality condition (\ref{eq-orthogonality_condition}) holds. Then there are infinitely many rank-$n$ geodesic arcs $\gamma_\geo^{1 \to 2}$ joining the non-faithful states $\rho_1$ and $\rho_2$,  which are given by (\ref{eq-Bures_geodesics_second_method}) below and have the same length 
	$\ell(\gamma_\geo)= d_{\Bures} (\rho_1,\rho_2)$. If 
	$r_1+r_2 =n$, the extensions of these arcs intersect $\partial \states$ only at $\rho_1$ and $\rho_2$. If 
	$r_1+r_2 > n$, they have $(q-2)$ other intersection states $\rho_k$, $k=3,\ldots,q$, where $q-2 > r_1+r_2-n$ is  
	the number of distinct nonzero eigenvalues of the operator
	\begin{equation} \label{eq-M_12}
	M_{\rho_1\rho_2}= \Pi_1 \rho_1^{-1/2} \Lambda_{\rho_2\rho_1} \rho_1^{-1/2}\, \Pi_1 \;,\;
	\Lambda_{\rho_2\rho_1} = | \sqrt{\rho_2} \sqrt{\rho_1} |\;.
	\end{equation}
	Moreover, 
	the states $\rho_k$ have fixed supports $\supp (\rho_k) = (\identity- P_k )\Hh$, where $P_k$ are the eigenprojectors  of $M_{\rho_1\rho_2}$ with nonzero eigenvalues.
}

\vspace{2mm}

Assuming that  (\ref{eq-restricted_orthogonality_cond}) and (\ref{eq-non_inculsion_condition}) hold, we can apply Theorem 2 to 
the manifold of quantum states $\statesof{\Hh_{12}}^\inv$
to conclude that there are infinitely many rank-$n_{12}$ geodesic arcs
$\gamma_\geo^{1 \to 2}$ with support $\Hh_{12}$ and length $ d_{\Bures} ( \rho_1, \rho_2)$. In fact, (\ref{eq-non_inculsion_condition}) 
entails $\Hh_{i} \not= \Hh_{12}$  for $i=1,2$, that is,
$r_1, r_2 < n_{12}$, implying that both $\rho_1$ and $\rho_2$ are on the boundary of $\statesof{\Hh_{12}}$. 
Note that one can not deduce from Theorem 2 the existence of geodesics $\gamma_\geo^{1 \to 2}$ of rank $k<n_{12}$.

\vspace{2mm}

Theorem 2 is proved in Appendix~\ref{sec-proof_theorem1}. 
Note the similarity of the statement about the intersections of $\gamma_\geo$ with  $\partial \states$ with what happens for geodesics joining invertible states $\rho,\sigma$ (see Sec.~\ref{sec_background_on_Bures_geodesics}). While  $M_{\sigma \rho}$ in (\ref{eq-definition_M}) is invertible, the operator (\ref{eq-M_12}) has support 
\begin{equation} \label{eq-support_M_rho1_rho2}
\supp ( M_{\rho_1\rho_2}) = \Hh_{1} \cap \Hh_{2} 
\end{equation}	    
and rank $r_1+r_2 - n$. 
In fact, recall from Sec.~\ref{sec-regularization_method} that 
$\ker ( \Lambda_{\rho_2\rho_1})= \Hh_1^\bot \oplus S_{21}$.
By the orthogonality hypothesis (\ref{eq-orthogonality_condition}) one has $\Hh_2^\bot \subseteq \Hh_1$. 
Thus $S_{21} = \rho_1^{-1/2} \Hh_2^\bot$, see (\ref{eq-definition_S_12}), and
\begin{eqnarray}
\nn
\ker ( M_{\rho_1\rho_2}) & = & 
\Hh_1^\bot \oplus \sqrt{\rho_1} \ker ( \Lambda_{\rho_2\rho_1})
\\
& = &  \Hh_1^\bot \oplus \sqrt{\rho_1} S_{21} = \Hh_1^\bot \oplus \Hh_2^\bot\;.
\end{eqnarray}
Hence $M_{\rho_1\rho_2}$ has rank $r_1+r_2-n$. Eq. (\ref{eq-support_M_rho1_rho2}) follows from the self-adjointness of 
 $M_{\rho_1\rho_2}$ and the relation $\Hh_1 \cap \Hh_2 = (\Hh_1^\bot \oplus \Hh_2^\bot)^\bot$.
Note that $M_{\rho_1\rho_2}=0$ when $r_1+r_2 = n$. In that case (\ref{eq-orthogonality_condition}) entails $\Hh_1^\bot = \Hh_2$ and $\Hh_2^\bot = \Hh_1$.

The explicit form of the geodesics $\rho_1\to \rho_2$
is derived in Appendix~\ref{sec-proof_theorem1}. It reads
\begin{eqnarray} \label{eq-Bures_geodesics_second_method}
\nonumber
&& 
\gamma_{\geo}^{1 \to 2} ( \tau) =\frac{1}{\sin^2 \tau_2}
\bigg( \sin^2 ( \tau_2 - \tau)\,  \rho_1  + \sin^2 ( \tau )\, \rho_2
\\ \nn
& & 
+\sin ( \tau_2 - \tau)  \sin ( \tau) \Big( 
\sqrt{\rho_2} \,U_{\rho_2\rho_1}^{(0)} \sqrt{\rho_1} + \sqrt{\rho_1} \,U_{\rho_2\rho_1}^{(0) \dagger} \sqrt{\rho_2}
\\
& & 
+\sin \tau_2 \, \dot{\rho}^{(12)}_1
\Big)
\bigg)\;,
\end{eqnarray}
where $\tau_2 = d_{\Bures}( \rho_1,\rho_2)$ is the arccos distance between the two states, $U_{\rho_2\rho_1}^{(0)}$ is the partial isometry  $\supp(\Lambda_{\rho_2\rho_1}) \to \supp (\Lambda_{\rho_1\rho_2})$ defined by the polar decomposition (\ref{eq-polar_decomp_non_invertible_states}), 
 and 
$\dot{\rho}^{(12)}_1$ is an arbitrary self-adjoint operator with support
$\Hh_1^\bot \oplus  \Hh_2^\bot$ 
satisfying the condition (\ref{eq-CNS_rho_0>0}) in Appendix~\ref{sec-proof_theorem1}.
Identifying 
$\dot{\rho}_1^{(12)}$ with the regularization-dependent term in (\ref{eq-Bures_geodesics_reg_bis}), namely,
\begin{equation} \label{eq-identification_tengent_op_two_methods}
\sin(\tau_2) \dot{\rho}_1^{(12)} = \sqrt{\rho_2}\,R_{\rho_2\rho_1} \sqrt{\rho_1}
+ {\mathrm{h.c.}}\;,
\end{equation}
the geodesics (\ref{eq-Bures_geodesics_second_method}) agree with those determined in Sec.~\ref{sec-regularization_method} by the regularization method. 
Note that  the operator sum in the \RHS of (\ref{eq-identification_tengent_op_two_methods})
has support in $\Hh_1^\bot \oplus \Hh_2^\bot$. Actually, since 
$\sqrt{\rho_1} \, S_{21} \subseteq \Hh_2^\bot \subseteq \Hh_1$, one has 
$\Pi_2^\bot \sqrt{\rho_1} \,\Pi_{S_{21}}=\sqrt{\rho_1}\, \Pi_{S_{21}}$, \ie,
$\Pi_{S_{21}} \sqrt{\rho_1}= \Pi_{S_{21}} \sqrt{\rho_1} \,\Pi_2^\bot$. 
Similar identities hold by exchanging $1$ and $2$. The result then follows from 
(\ref{eq-def_R_rho2_rho1}).

\vspace{1mm}	
	
Let us make here a side comment. Let us introduce the projective measurement $\{P_j\}_{j=1}^q$,
with $P_i = \Pi_i^\bot$ for $i=1,2$ and $P_k$ the eigenprojectors of $M_{\rho_1\rho_2}$ with positive eigenvalues. It is shown in Appendix~\ref{sec-proof_theorem1} that $\{P_j\}_{j=1}^q$  is an optimal measurement for the two following discrimination tasks: (1)  unambiguously discriminating $\rho_1$ and $\rho_2$ with the smallest possible inconclusive outcome probabilities $p_{?|1}=1-p_{2|1}$ and $p_{?|2}= 1 - p_{1|2}$; (2) 
distinguishing $\rho_1$ and $\rho_2$ from the statistics of the measurement data.
Task (1) is easy to solve under the orthogonality hypothesis (\ref{eq-orthogonality_condition}), see e.g.~\cite{myreview}. Task (2), which amounts to maximize the classical Hellinger distance between the two measurement distributions $\{ p_{j|1}\}_{j=1}^n$ and $\{ p_{j|2}\}_{j=1}^n$, is much less trivial, since $\rho_1$ and $\rho_2$ are in general not orthogonal.

\vspace{1mm}

In order to prove consistency of the results of this section with those of Sec.~\ref{sec-regularization_method}, it is worth checking 
that conditions (\ref{eq-restricted_orthogonality_cond}) and (\ref{eq-non_inculsion_condition}) imply that condition (\ref{eq-sufcond1}) of Theorem~1 
  is not fulfilled. With this aim, let us show that
(\ref{eq-restricted_orthogonality_cond}) implies
\begin{equation} \label{eq-equivalent_cond_orthogonality_kernels}
\Hh_1^\bot \cap \Hh_{12} \subseteq \Hh_1^\bot \cap \Hh_2 \;.
\end{equation}
Indeed, since 
$\Hh_1^\bot \cap \,\Hh_2^\bot \subseteq \Hh_2^\bot$ is the orthogonal complement of $\Hh_{12}=\Hh_1 \oplus \Hh_2$, one can decompose $\Hh_2^\bot$ as
\begin{equation} \label{eq-decomp_kernel_into_orthogonal_subspaces}
\Hh_2^\bot = \Hh_2^\bot \cap \,\Hh_{12} \, \oplus \, \Hh_1^\bot \cap\,\Hh_2^\bot\;.
\end{equation}
If (\ref{eq-restricted_orthogonality_cond}) holds then  $\Hh_1^\bot \cap\,\Hh_{12}$ is orthogonal to the first subspace in the \RHS.
It is also orthogonal to the second subspace  
because $\Hh_{12} \perp \Hh_1^\bot \cap \,\Hh_2^\bot$.   
Thus $\Hh_1^\bot \cap \Hh_{12} \,\perp \, \Hh_2^\perp$, which
implies (\ref{eq-equivalent_cond_orthogonality_kernels}).
Now, we have shown above that (\ref{eq-non_inculsion_condition}) entails $r_1, r_2 < n_{12}$. Hence 
the subspace in the \LHS of (\ref{eq-equivalent_cond_orthogonality_kernels}) 
is non-trivial. Thus $S_{12}\not= \{ 0\}$.
The same is true for $S_{21}$ thanks to the symmetry of (\ref{eq-restricted_orthogonality_cond}) with respect to the exchange of $1$ and $2$. In summary, 
\begin{equation} \label{eq-implication_orthogonality_triviality_of_S}
\text{(\ref{eq-restricted_orthogonality_cond}) and (\ref{eq-non_inculsion_condition})} 
\;\Rightarrow\; 
S_{12}\not= \{0\} \;,\; S_{21} \not= \{0\}\;.
\end{equation}
The converse implication is not true except when $r_1 =n-1$ or $r_2=n-1$.
In the latter case $\dim \Hh_2^\bot=1$ and thus
$S_{21}$ is non-trivial if and only if  $\Hh_2^\bot \subset \Hh_1$, which is equivalent to $\Hh_1^\bot \perp \Hh_2^\bot$; when $r_1=n-1$ a similar statement holds with $S_{12}$.
\vspace{1mm}


Note that if $\rho_1 = \ketbra{\psi_1}{\psi_1}$ and $\rho_2 = \ketbra{\psi_2}{\psi_2}$ are pure states, one has $\Hh_{12} = \Span \{ \ket{\psi_1}, \ket{\psi_2}\}$ and 
(\ref{eq-restricted_orthogonality_cond}) is equivalent to $\braket{\psi_1}{\psi_2}=0$. Thus Theorem~2 gives the same result  as Theorem~1 (see Sec.~\ref{sec-regularization_method}): geodesics joining non-orthogonal pure states are rank-1, and there are infinitely many rank-2 geodesics joining orthogonal pure states, having support $\Hh_{12}$.

\vspace{2mm}

 In conclusion, we deduce from Theorem~2 that:

\vspace{1mm} \noindent {\it 
	In order to have a unique geodesic $\gamma^{1 \to 2}_\geo$ it is necessary (but not sufficient) that either (\ref{eq-restricted_orthogonality_cond}) or (\ref{eq-non_inculsion_condition}) is not fulfilled. In the first case $\gamma_\geo^{1 \to 2}$ has a rank smaller than $n_{12}$ and in the second one $\gamma_\geo^{1\to 2}$ is unique and $\supp ( \gamma_\geo) = \Hh_{12}$.
}

\section{Boundary geodesics for qutrits} \label{sec-qutrit}

		\begin{table*}[htbp]
		\centering
		\renewcommand{\arraystretch}{1.25}
			\begin{tabular}{cccccc p{9.5cm}}
			\hline
			$(r_1,r_2)$  & $s_{12}$ & $s_{21}$  & $\Hh_1 \perp \Hh_2$ 
			&		  $\Hh_1^\bot \cap \Hh_{12}\perp \Hh_2^\bot \cap \Hh_{12}$ 
			& $n_{12}$
			&  {\textbf{Description of the shortest geodesics}} $\rho_1\to \rho_2$\\ 
			\hline\hline
			(1, 1) & 1 & 1 & yes & yes & 2
			& Infinitely many rank-$2$ geodesics and rank-$1$ (Fubini--Study) geodesics,  
			the rank-$2$ geodesics have support $\Hh_{12}=\Span \{ \ket{\psi_1},\ket{\psi_2}\}$ \\
			(1, 1) & 0 & 0  & no & no & 2
			& Unique rank-$1$ (Fubini--Study) geodesic \\
			\hline
			(1, 2) & 2 & 1 & yes & yes  & 3
			& Infinitely-many  rank-$3$ and rank-$2$ geodesics, the rank-$3$ geodesics intersect the boundary only at $\rho_1$ and $\rho_2$ \\
			(1, 2) & 1 & 0 & no & no (save when $\Hh_1 \subset \Hh_2$) & 3 (2)
			& Unique rank-$2$ geodesic with support $\Hh$ \\
			\hline
			(2, 2) & 1 & 1 & no & yes  & 3
			&    Infinitely many  rank-$3$ and rank-$2$ geodesics,  the rank-$3$ geodesics intersect the boundary at
			$\rho_1$, $\rho_2$ and another  state $\rho_3$ of rank $2$\\
			(2, 2) & 0 & 0  & no & no (save when $\Hh_1=\Hh_2$) & 3 (2)
			& Unique rank-$2$ geodesic with support $\Hh$ (save when $\Hh_1 = \Hh_2$, in which case $\supp ( \gamma_\geo)=\Hh_1= \Hh_2$) \\
			\hline
			(2, 3) & 1 & 0 &  no  & yes (since $\Hh_2^\bot = \{0\}$) &  3
			& Unique rank-$3$ geodesic  \\
			\hline
			(3, 3) & 0 & 0 & no  &  yes (since $\Hh_2^\bot = \{0\}$) &  3
			& Unique rank-$3$ geodesic  \\
			\hline
		\end{tabular}
		\caption{Classification of the geodesics for a qutrit ($n=3$) according to the ranks $r_1$ and $r_2$ of the end-states $\rho_1$ and $\rho_2$, the dimensions  $s_{12} = \dim(S_{12})=\dim(  \Hh_1^\perp \cap \Hh_2 )$ and  $s_{21}= \dim(S_{21})=\dim(\Hh_2^\perp \cap \Hh_1 )$
			and orthogonality conditions on their supports $\Hh_i$ and kernels $\Hh_i^\bot$. Here $\Hh_{12} =\Hh_1+\Hh_2$ and $n_{12} = \dim (\Hh_{12})$.}	
		\label{tab:geodesics_N3}
	\end{table*}

Let us apply our general results to the special case  of a qutrit ($n=3$).
We obtain the following classification of the qutrit geodesic arcs $\gamma_\geo^{1 \to 2}$ 
 according to the values $r_1$ and $r_2$ of the ranks of $\rho_1$ and $\rho_2$.
Without loss of generality one can assume that $r_1 \leq r_2$. In fact, 
$\gamma_\geo^{1 \to 2}$ is a geodesic arc from $\rho_1$ to $\rho_2$ if and only if
its time-reversal $(\gamma_\geo^{1 \to 2})_{\TR}$ is a geodesic arc from $\rho_2$ to $\rho_1$, with
$(\gamma_\geo^{1 \to 2})_{\TR}(\tau)= \gamma_\geo^{1 \to 2}(\theta - \tau)$. 
We distinguish the following cases.

\vspace{1mm}

\begin{enumerate}
	
\item[1)]  \label{case(1,1)} 
$r_1=r_2=1$: \\
As for higher dimensional spaces, $\gamma_\geo^{1\to 2}$ is unique  
and is a Fubini--Study geodesic save when $\rho_1$ and $\rho_2$  are orthogonal. In the latter case, there are infinitely many rank-$2$ (qubit-like) geodesics with support $\Hh_1 + \Hh_2$ and infinitely many rank-$1$ (Fubini--Study) geodesics. 	
	
\item[2)] \label{case(1,2)} $r_1=1$, $r_2=2$: \\
By the observation following Theorem~1, one has (\ref{eq-sufcond0}) $\Leftrightarrow \Hh_2^\bot \cap \Hh_1 = \{0\}$.
Setting
$\rho_1 = \ketbra{\psi_1}{\psi_1}$ and noting that the kernel $\Hh_2^\bot$ and  support $\Hh_1 = \complex \ket{\psi_1}$ are one-dimensional, one has
$\Hh_2^\bot \cap \Hh_1 \not= \{0\}$ if and only if $\ket{\psi_1} \perp  \Hh_2$, \ie, $\Hh_1 = \Hh_2^\bot$.
In all cases $\Hh_1^\bot \cap \, \Hh_2\not=\{0\}$. 
Therefore

\vspace{1mm}

\begin{enumerate}

\item[2a)] if $\ket{\psi_1}\not\perp \Hh_2$ then $\Hh_2^\bot \cap \Hh_1 = \{0\}$ and condition (\ref{eq-sufcond0}) of Theorem~1 holds, implying that there is a unique $\gamma_\geo^{1\to 2}$. This geodesic is of rank-$2$ since (\ref{eq-orthogonality_condition}) is not fulfilled.
	
\item[2b)] if $\ket{\psi_1} \perp \Hh_2$ then  $\Hh_2^\bot \cap \Hh_1 \not= \{0\}$,	condition (\ref{eq-sufcond1}) does not hold and $\Hh_1^\bot \perp \Hh_2^\bot$, implying that there are infinitely many rank-$3$ geodesics.
We show below that there are also infinitely many rank-$2$ geodesics. 

\end{enumerate}

\item[3)]  \label{case(2,2)} $r_1=r_2=2$: \\
By the observation following Theorem~1, $S_{21} \not= \{ 0 \}$
is equivalent to $S_{12} \not= \{0\}$ and also equivalent to the orthogonality of the one-dimensional kernels $\Hh_1^\bot$ and $\Hh_2^\bot$. Therefore:

\begin{enumerate}
\item[3a)] If $\Hh_1^\bot \not\perp \Hh_2^\bot$ there is a unique $\gamma_\geo^{1\to 2}$, which is rank-$2$.

\item[3b)] If $\Hh_1^\bot \perp \Hh_2^\bot$  there are infinitely many full-rank geodesics (which intersect the boundary at a third state $\rho_3$ of rank $2$ and kernel $\Hh_3^\bot \perp \Hh_1^\bot \oplus \Hh_2^\bot$). One finds numerically that there also exists infinitely many rank-$2$ geodesics. 
	
\end{enumerate}
   
\item[4)]  \label{case(2,3)} $r_1=2$, $r_2=3$:\\
Then, according to the general results on bulk geodesics  (see Sec.~\ref{sec_background_on_Bures_geodesics}), there is a unique shortest geodesic arc joining $\rho_1$ and $\rho_2$, given 
 by the time reversal of  (\ref{eq-Bures_geodesics_V}) with $(\rho,\sigma)=(\rho_2,\rho_1)$.

\item[5)]  \label{case(3,3)} $r_1=r_2=3$:\\
Same as in case 4).

\end{enumerate}

Our conclusions are summarized in Table~\ref{tab:geodesics_N3}.	
As in the qubit case, for a qutrit the necessary and sufficient condition for uniqueness of $\gamma_\geo^{1\to 2}$ 
is  the non-orthogonality of $\Hh_1^\bot = \ker(\rho_1)$ and $\Hh_2^\bot = \ker(\rho_2)$. We will see in the next section that this statement is wrong for dimensions  $n \geq 4$. 

\vspace{2mm}

It is instructive to work out the explicit form of the geodesics in case 2b). Choosing the orthonormal basis $\{ \ket{1},\ket{2},\ket{3}\}$ of $\complex^3$ such that 
$\rho_1 = \ketbra{1}{1} = \Pi_1$ and $\Hh_2= \Span \{ \ket{2},\ket{3}\}$, since $\Hh_1^\bot = \Hh_2$ one has $S_{12} =  \Hh_2$ and $S_{21} = \Hh_1$. Furthermore, the states $\rho_1$ and $\rho_2$ being orthogonal, one has $\theta=\pi/2$ and 
$\Lambda_{\rho_2\rho_1} = \Lambda_{\rho_1\rho_2}=0$. Hence by (\ref{eq-decomp_U_21^reg}) the unitary $U_{\rho_2\rho_1}^\reg$ reduces to its regularization-dependent part $U_{\rho_2\rho_1}^\bot$, which is an arbitrary $3\times 3$ unitary matrix.
Let us set
\begin{equation}
\ket{\widetilde{\phi}}= U_{\rho_2\rho_1}^\bot \ket{1} = \alpha_2 \ket{2} + \alpha_3 \ket{3}\;,
\end{equation}
which satisfies $\| \widetilde{\phi}\|^2  \leq 1$.   Without loss of generality one can assume 
that $\rho_2 = p_2 \ketbra{2}{2} + (1-p_2) \ketbra{3}{3}$.
Eq.~(\ref{eq-Bures_geodesics_reg_bis} gives the following matrix of  $\gamma_\geo^{1\to 2} (\tau)$ in the $\{\ket{1},\ket{2},\ket{3}\}$-basis:  
\begin{equation}
\begin{pmatrix}
\cos^2 \tau & \sqrt{p_2} \,\alpha_2^\ast \frac{\sin (2\tau )}{2} & 
\sqrt{1-p_2}\, \alpha_3^\ast \frac{\sin(2 \tau)}{2} 
\\[1mm]
\sqrt{p_2} \,\alpha_2  \frac{\sin (2\tau)}{2} & 
p_2 \sin^2 \tau  &  0
\\[1mm]
\sqrt{1-p_2} \,\alpha_3  \frac{\sin (2\tau)}{2} & 
 0 &  (1-p_2) \sin^2 \tau   
\end{pmatrix}
\;.
\end{equation}
The complex coefficients $\alpha_2$ and $\alpha_3$ in this matrix are arbitrary,
with
$|\alpha_2|^2+ |\alpha_3|^2 \leq 1$.
Using $\det (\gamma_\geo^{1\to 2} (\tau))= p_2 (1-p_2) \sin^4\tau \cos^2 \tau 
( 1 - \| \widetilde{\phi}\|^2)$, one deduces that the geodesic is rank-3 if 
$|\alpha_2|^2+ |\alpha_3|^2 < 1$ and rank-2 if $|\alpha_2|^2+ |\alpha_3|^2 = 1$.

\begin{table*}[t]
\centering
\scriptsize
\setlength{\tabcolsep}{3pt}
\renewcommand{\arraystretch}{1.1}
	\begin{tabular}{c c c c c c p{9.5cm}}
			\toprule
			$(r_1,r_2)$ & $s_{12}$ & $s_{21}$  & $\Hh_1\perp \Hh_2$ &  $\Hh_1^\bot \cap \Hh_{12}\perp \Hh_2^\bot \cap \Hh_{12}$  & $n_{12}$ 
			&  {\textbf{Description of the shortest geodesics}} $\rho_1\to \rho_2$ \\
			\midrule
			$(1,1)$ & $1$ & $1$ & yes & yes &   2
			& Infinitely many rank-$2$  and rank-$1$ geodesics. \\			
			$(1,1)$ & $0$ & $0$ & no & no &  2
			& Unique rank-$1$ (Fubini--Study) geodesic. \\ \hline
			$(1,2)$ & $2$ & $1$ & yes & yes &  3
			& Infinitely many rank-$3$ and rank-$2$ geodesics. No rank-4 geodesics. \\
			$(1,2)$ & $1$ & $0$ & no & no (save when $\Hh_1 \subseteq \Hh_2$) & 3 (2)
			& Unique geodesic, with rank $2$ and support $\Hh_{12}$. \\ \hline
			$(1,3)$ & $3$ & $1$ & yes & yes & 4 
			& Infinitely many rank-$4$ and rank-$3$  geodesics. The rank-$4$ geodesics intersect the boundary only at $\rho_1$ and $\rho_2$.\\
			$(1,3)$ & $2$ & $0$ & no & no (save when $\Hh_1 \subseteq \Hh_2$) & 4 (3) 
			& Unique geodesic, with rank $3$ and support $\Hh_{12}$. \\ \hline
			$(2,2)$ & $2$ & $2$ & yes & yes & 4 
			& Infinitely many rank-$4$, rank-$3$ and rank-$2$ geodesics. The rank-$4$ geodesics  intersect the boundary only at $\rho_1$ and $\rho_2$. \\
			$(2,2)$ & $1$ & $1$ & no & no &  4
			& Infinitely many rank-$3$ and rank-$2$ geodesics. No rank-$4$ geodesics. \\
			        &     &     & no & yes & 3 
			& Infinitely many rank-$3$ and rank-$2$ geodesics. No rank-$4$ geodesics. \\
			$(2,2)$ & $0$ & $0$ & no & no (save when $\Hh_1=\Hh_2$) & 4, 3, (2)
			&  Unique geodesic, with rank $2$ and support $\Hh_{12}$. \\ \hline
			$(2,3)$ & $2$ & $1$ & no & yes & 4 
			& Infinitely many rank-$3$ and rank-$4$ geodesics. The rank-$4$ geodesics have three boundary intersections at $\rho_1,\rho_2$ and some other rank-$3$ state.  \\
			$(2,3)$ & $1$ & $0$ & no & no (save when $\Hh_1 \subseteq \Hh_2$) & 4 (3) 
			& Unique  geodesic, with rank $3$ and support $\Hh$ (save when $\Hh_1 \subseteq \Hh_2$). This geodesic intersects the rank-$2$ stratum at $\rho_1$ and another $\rho_3$ state of rank $2$. \\ \hline
			$(3,3)$ & $1$ & $1$ & no & yes & 4
			& Infinitely many  rank-$3$ and rank-$4$ geodesics. The rank-$3$ geodesics has one intersection with the rank-$1$ stratum, while the rank-$4$ geodesics intersect the boundary at $\rho_1$, $\rho_2$ and at another  state $\rho_3$ of rank $2$. \\
			$(3,3)$ & $0$ & $0$ & no & no (save when $\Hh_1=\Hh_2$) & 4 (3) 
			& Unique geodesic, with rank $3$ and support $\Hh$ (save when $\Hh_1=\Hh_2$). \\
			\bottomrule
		\end{tabular}
		
		\caption{Same as in Table~\ref{tab:geodesics_N3} for a two-qubit system ($n=4$). Note that in the case $(r_1,r_2)=(2,2)$, $s_{12}=s_{21}=1$ and $n_{12}=4$, one has infinitely-many geodesics even though the orthogonality condition on 
		the restriction  to $\Hh_{12}$ of the kernels of $\rho_1$ and $\rho_2$  is not fulfilled (see the example in Sec.~\ref{sec-2qubit}).
		}
		\label{tab:classification_geodesics}
\end{table*}

\section{Boundary geodesics for two qubits} \label{sec-2qubit}

In this section we focus on a two-qubit system ($n=4$). 
The classification of the geodesics arcs $\gamma_\geo^{1 \to 2}$ 
according to the values $r_1$ and $r_2$ of the ranks of $\rho_1$ and $\rho_2$
is given in Table~\ref{tab:classification_geodesics}. 
We assume without loss of generality that $1 \leq r_1 \leq r_2 < 4$.
The values of $s_{12}$ and $s_{21}$ in the second and third columns follow from
\begin{equation}
0 \leq s_{21} = s_{12} - (r_2 - r_1) \leq s_{12} \leq \min \{ n-r_1,r_2\}\;,
\end{equation}
where we have used (\ref{eq-relation_dimensions_S}) and $s_{12} = \dim (\Hh_1^\bot \cap \Hh_2)$.
The orthogonality conditions in the fourth and fifth columns are obtained from the following observations: (i) If either $r_1=1$ or $r_1=r_2= s_{12}=s_{21}$, then
\begin{eqnarray} \label{eq-orthogonality_cond_r1=1} 
s_{21} \not= 0 & \;\;\Leftrightarrow\;\; & \Hh_1 \subseteq \Hh_2^\bot \;\Leftrightarrow \; \Hh_2 \subseteq \Hh_1^\bot
\\ \nn
& \Rightarrow & \Hh_1 = \Hh_2^\bot \cap \Hh_{12} \, \perp \, \Hh_2 = \Hh_1^\bot \cap \Hh_{12}\;.	
\end{eqnarray} 
(ii) If $r_2 = n-1$
then 
\begin{eqnarray} \label{eq-orthogonality_cond_r2=n-1} 
	s_{21} \not= 0 & \;\;\Leftrightarrow\;\; &  \Hh_2^\bot \subseteq \Hh_1 \;\Leftrightarrow \; \Hh_1^\bot \subseteq \Hh_2
	\\ \nn
	& \Rightarrow & \Hh_2^\bot \, \perp \, \Hh_1^\bot \;\text{ and } \; \Hh_{12} = \Hh\; .	
\end{eqnarray} 
The statements in the last column about the uniqueness and rank of $\gamma_\geo^{1\to2}$ are deduced from the results of Secs.~\ref{sec-regularization_method} and \ref{sec-extension_method}. 
The existence of geodesics with lower ranks is inferred by assuming that the operator $R_{\rho_2\rho_1}$ in (\ref{eq-Bures_geodesics_reg_bis}) can be an arbitrary contraction and determining the ranks of 
$\gamma_\geo^{1 \to 2}(\tau)$ numerically, as illustrated in the example below.

\vspace{1mm}

Let us consider a specific example of 
two density operators $\rho_1,\rho_2$ with rank $r_1=r_2=2$ given by a diagonal state and a Werner state,
\begin{align}
\nn
\rho_1 &= p \ketbra{10}{10} + (1-p) \ketbra{11}{11}
\\
\rho_2 & = (1-q) \ketbra{00}{00} + q \ketbra{\Psi_+}{\Psi_+}\; , 
\label{eq:rho2_ex}
\end{align}
where  $0\leq p,q \leq 1$, $\{|00\rangle,|01\rangle,|10\rangle,|11\rangle\}$ is the computational basis of $\complex^4$, and $\ket{\Psi_{\pm}}$ are the Bell states,
\begin{equation}
|\Psi_{\pm}\rangle = \frac{|01\rangle \pm  |10\rangle}{\sqrt{2}}\;.
\end{equation}
One easily finds
\begin{equation}
\Hh_1^\bot  =\Span \{ \ket{00},\ket{01}\}\; , \; \Hh_2^\bot = \Span \{ \ket{\Psi_-}, \ket{11}\}
\end{equation}
and
\begin{equation}
S_{12}=\Hh_1^\bot \cap \,\Hh_2 = \complex \ket{00} \; ,\;
S_{21}= \Hh_2^\bot \cap \,\Hh_1 = \complex \ket{11}\;.
\end{equation} 
Hence $s_{12}=s_{21}=1$. Moreover $\Hh_{12}=\complex^4$, so that $n_{12}=4$.
Note that condition (\ref{eq-sufcond0}) is not fulfilled, even though  the orthogonality condition (\ref{eq-restricted_orthogonality_cond}) (or, equivalently, (\ref{eq-orthogonality_condition})) does not hold.
This gives a counterexample illustrating that
the converse of (\ref{eq-implication_orthogonality_triviality_of_S}) is not valid. 
Then Theorem~2 only excludes rank-$4$ geodesics. One infers from our conjecture in Sec~\ref{sec-regularization_method} that there are infinitely many geodesics $\gamma_\geo^{1 \to 2}$.

A direct computation gives
$\cos\theta = \sqrt{F(\rho_1,\rho_2)} = \sqrt{ pq /2}$ and
\begin{equation}
U_{\rho_2\rho_1}^{(0)}  =  \ketbra{\Psi_+}{10}
\;\; , \;\;  R_{\rho_2\rho_1 } = z \ketbra{00}{11} \;, 
\end{equation}
where the  complex parameter $z=\bra{00} U_{\rho_2\rho_1}^\bot \ket{11}$  satisfies $0\le |z|\le 1$.
The Bures geodesics (\ref{eq-Bures_geodesics_reg_bis}) connecting $\rho_1$ and $\rho_2$ can be written explicitly as a one--parameter family 
\begin{equation}
\begin{aligned}
&	\gamma_\geo^{1\to2}(\tau)
	=\frac{1}{\sin^2\theta}\bigg(
	\sin^2(\theta-\tau)\,\rho_1
	+\sin^2 (\tau  ) \rho_2
	\\
	&\quad
	+\sin (\tau )\sin(\theta-\tau ) \Big( \sqrt{pq}\,
	|\Psi_+\rangle\langle 10|
	\\
	&\quad
	+z \sqrt{(1-p)(1-q)}
	|00 \rangle\langle 11| + {\mathrm{h.c.}}\Big)
	\Bigg)\;.
	\end{aligned}
\label{eq:rho_g_Z}
\end{equation}
Computing numerically the spectrum of $\gamma_\geo^{1\to 2}(\tau)$ as a function of $\tau$, one finds that (see Fig.~\ref{fig:spectrum_a})
 \begin{itemize}
 		\item when $0<|z|<1$, the geodesic $\gamma_\geo^{1\to 2}$ has rank $3$;
 		\item when $|z|=1$, the geodesic $\gamma_\geo^{1\to 2}$ has rank $2$.
 \end{itemize}

\begin{figure}[t]
	\centering
	\includegraphics[width=7.5cm]{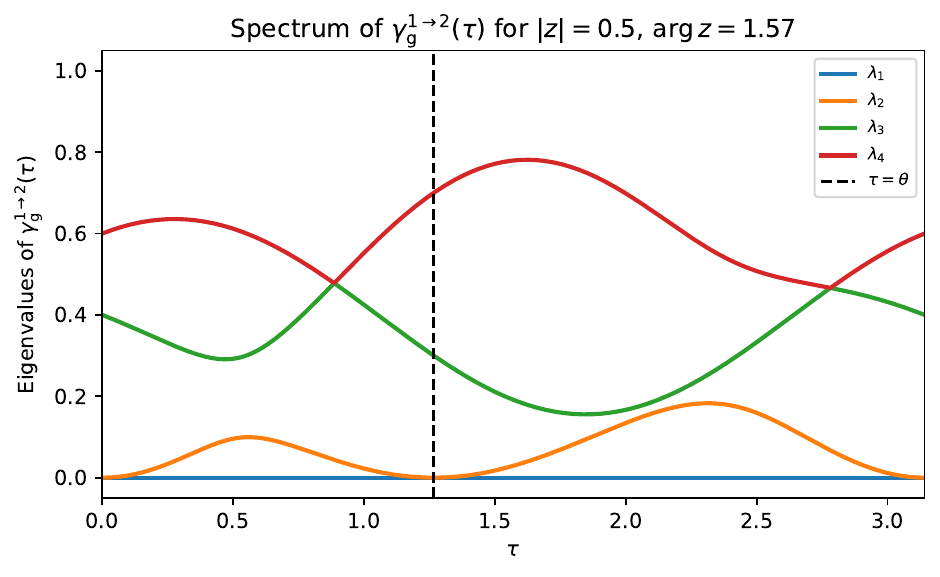}
	\includegraphics[width=7.5cm]{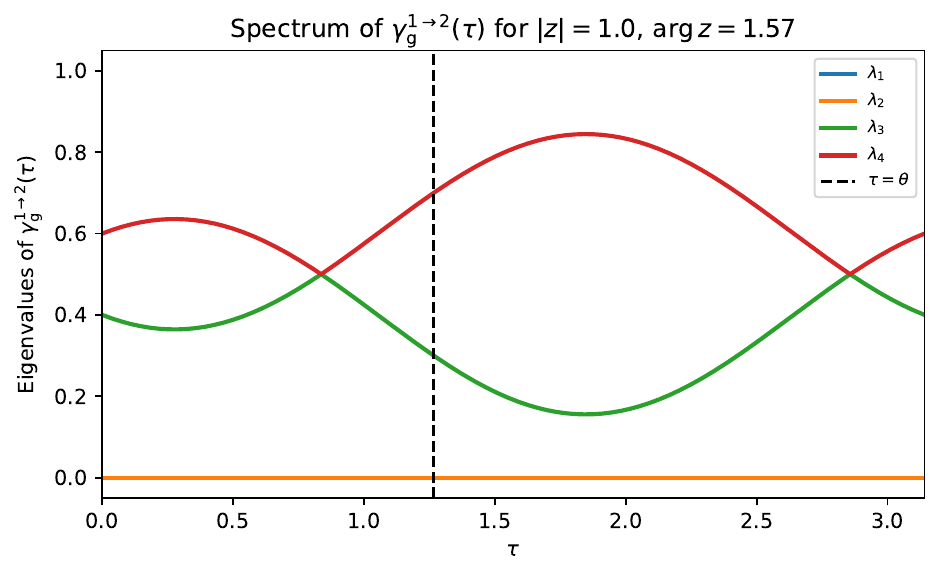}
	\caption{Spectrum of $\gamma_\geo^{1 \to 2}(\tau)$ vs time $\tau$ for $|z|=0.5$ and $|z|=1.0$, showing that  for all $\tau \in (0,\theta)$, $\gamma_\geo^{1\to 2}(\tau)$ has rank $3$ in the first case and rank $2$ in the second case.
		The vertical dashed line indicates $t=\theta$ and $\arg (z) = \pi$.}
	\label{fig:spectrum_a}
\end{figure}

\section{Conclusions and perspectives} \label{sec-conclusion}

In this work we have determined the Bures geodesics between
non-faithful quantum states. This subtlety of this problem is related to the fact that 
the set $\states$ of
density matrices is not globally a smooth manifold, but rather a stratified
manifold whose lower-rank strata possess a nontrivial differential structure
\cite{DAndrea2021}. While Bures geodesics are well understood in the interior
$\states^\inv$
of this manifold through the purification bundle and the theory of
Riemannian submersions, their extension to boundary strata requires additional
care.

We developed two complementary approaches to tackle this problem. The first one is
based on regularizing non-faithful states by faithful ones, constructing the
corresponding geodesics in $\states^\inv$, and then taking the limit. When this
limit is independent of the chosen regularization, a unique shortest 
geodesic arc is obtained. This approach led us to identify a geometric condition
that guarantees uniqueness. When this condition fails, the limiting procedure
yields infinitely many geodesic arcs having the same length. Some aspects of this approach remain open, in particular
the characterization of the dependence of the rank of the geodesics on
the regularization curves.
The second approach consists in applying known results for geodesics in  $\states^\inv$ to the subspace spanned by the
supports of the two states. This yields the
family of geodesics when their kernels restricted to this subspace are orthogonal. This construction provides a direct geometric criterion for the
existence of geodesics of maximal rank and is fully consistent with the
regularization method. 

In the particular case of pure end-states, our results show that 
for non-orthogonal states there is a unique shortest geodesic arc connecting them, which is a rank-$1$ (Fubini-Study) geodesic, while when they are orthogonal there are infinitely many arcs, having either rank $1$ or rank $2$. This result has important implications for the Quantum Speed Limit (QSL). As far as we are aware, it was not known previously, save in the qubit case.

More generally, our results suggest that the uniqueness of Bures
geodesics between non-faithful states is primarily controlled by the relative
geometry of their supports and kernels, rather than by algebraic properties
such as commutativity.

These results have direct implications for the QSL. As shown in Sec.~\ref{sec_QSL},
Bures
geodesics determine evolutions that saturate the Mandelstam-Tamm bound under the constraint of a fixed energy cost. Therefore, their non-uniqueness
implies the existence of infinitely many optimal evolutions
with the same minimal time. These evolutions may have different ranks, offering
some freedom for selecting the most convenient implementation. Beyond
QSL, the geodesics constructed in this paper may also be useful in
quantum metrology and optimal quantum control, where Bures-geodesic trajectories
provide natural candidates for optimal state transformations~\cite{moi_geodesics_QMetrology,Mauro-thesis,Giovannetti2009,Deffner2014}.

An increasing interest in the geometric structure of classical probability space and quantum state space has emerged in the last two decades in information geometry
\cite{Amari_book00,Amari_book16}. In this context, the determination of
geodesics is a central problem, since geodesics encode optimal paths,
distances, and distinguishability properties between states.

\vspace{0.3cm}
\noindent{\bf\large Acknowledgments} SC acknowledges support from Chilean ANID through Fondecyt Postdoctoral Grant No.~3240514.

\vspace{0.3cm}

\newpage

\appendix
\renewcommand{\theequation}{\Alph{section}\arabic{equation}}
\setcounter{equation}{0}

\section{Proof that $U_{\rho_2\rho_1}^{(0)}$ is regularization-independent when $\Lambda_{\rho_2\rho_1}$ has degenerate positive eigenvalues} \label{sec-Lambda_21_degenerate_eigenvalues}

In this appendix we relax the assumption made in Sec.~\ref{sec-regularization_method}
that $\Lambda_{\rho_2\rho_1}$ has non-degenerate positive eigenvalues. We show that even when this assumption is not fulfilled, the partial isometry
$U_{\rho_2\rho_1}^{(0)}$ in the regularized unitary (\ref{eq-decomp_U_21^reg})
is regularization-independent and given by the polar decomposition (\ref{eq-polar_decomp_non_invertible_states}).

We assume that the curves $\eps_i \mapsto \rho_i(\eps_i)$ lift  
the degeneracies of all eigenvalues of the unperturbed operator, \ie,  $\Lambda_{21}(\veceps)$ has non-degenerate spectrum when $\eps_1,\eps_2>0$.
Then
the eigenvectors  $\ket{u_k (\veceps)}$ and $\ket{v_k (\veceps)}$ of $\Lambda_{21}^2(\veceps)$ and $\Lambda_{12}^2(\veceps)$ 
with the same eigenvalue $\lambda_k^2(\veceps)>0$ are related  by 
\begin{equation} \label{eq-relation_eigenvectors_Lambda_21_Lambda_12}
\ket{v_k (\veceps)} = \lambda_k(\veceps)^{-1} \sqrt{\regrhotwo} \sqrt{\regrhoone} \,\ket{u_k (\veceps)}
\end{equation}
(we use here the notation of Sec.~\ref{sec-regularization_method}).
Indeed, if $\Lambda_{21}^2(\veceps) \ket{u_k (\veceps)} = \lambda_k^2(\veceps) \ket{u_k (\veceps)}$, then by right-multiplying this equation by 
$\sqrt{\regrhotwo}\sqrt{\regrhoone}$ one gets
\begin{align}
\nn
& \Lambda_{12}^2(\veceps) \sqrt{\regrhotwo}\sqrt{\regrhoone} \ket{u_k (\veceps)}
\\
& \qquad = \lambda_k^2(\veceps)  \sqrt{\regrhotwo}\sqrt{\regrhoone} \ket{u_k (\veceps)}\;.
\end{align}
By our non-degeneracy assumption, it follows that 
the eigenvectors of $\Lambda_{12}^2(\veceps)$ with eigenvalue 
$\lambda_k(\veceps)$ are proportional to $ \sqrt{\regrhotwo}\sqrt{\regrhoone} \ket{u_k (\veceps)}$, which yields (\ref{eq-relation_eigenvectors_Lambda_21_Lambda_12}) upon normalization.
Taking the limit $\eps_1,\eps_2\to 0$ in this equation  and assuming $\lambda_k(\veceps) \to \lambda_k >0$, one obtains
\begin{equation} \label{eq-relation_between_v_and_u}
\ket{v_k^{(0)}} 
= \lambda_k^{-1} \sqrt{\rho_2} \sqrt{\rho_1} \ket{u_k^{(0)}}\;.
\end{equation}
This identity implies that if $\ket{u_k^{(0)}}$ belongs to the eigenspace of 
$\Lambda_{\rho_2\rho_1}^2$ with eigenvalue $\lambda_k$ then $\ket{v_k^{(0)}}$ belongs to the eigenspace of 
$\Lambda_{\rho_1\rho_2}^2$ with the same eigenvalue. If $\lambda_k$ is degenerate,
the reciprocal implication is, however, not true. This means that (\ref{eq-relation_between_v_and_u}) sets an extra constraint, which is not contained in 
condition (\ref{eq-polar_dec_bis}) for $\eps_1=\eps_2=0$.

Recall that $U_{\rho_2\rho_1}^{(0)} $ is defined as
\begin{equation} \label{eq-regularization_indep_part_of_unitary}
U_{\rho_2\rho_1}^{(0)} = \Pi_{\rho_1\rho_2} U_{\rho_2\rho_1}^\reg  \Pi_{\rho_2\rho_1} = \sum_{k=1}^m \ketbra{v_k^{(0)}}{u_k^{(0)}}\;.
\end{equation}
Plugging (\ref{eq-relation_between_v_and_u}) into that equation we get
\begin{eqnarray}
\nn
U_{\rho_2\rho_1}^{(0)} \Lambda_{\rho_2\rho_1} \ket{u_k^{(0)}}
& = & 
\lambda_k U_{\rho_2\rho_1}^{(0)} \ket{u_k^{(0)}}=\lambda_k \ket{v_k^{(0)}} 
\\
& = & 
\sqrt{\rho_1} \sqrt{\rho_2} \ket{u_k^{(0)}}
\end{eqnarray}
for any $k=1,\ldots,m$. This implies that $U_{\rho_2\rho_1}^{(0)}$ satisfies (\ref{eq-polar_decomp_non_invertible_states}), \ie, it is the unique partial isometry $\supp (\Lambda_{\rho_2\rho_1}) \to \supp (\Lambda_{\rho_1\rho_2})$ in the polar decomposition of $\sqrt{\rho_2}\sqrt{\rho_1}$.
Right-multiplying both members of this equation by $\Lambda_{\rho_2\rho_1}^{-1}$, one concludes that
$U_{\rho_2\rho_1}^{(0)} $ only depends on $\rho_1$ and $\rho_2$.

\section{Contractions $R_{\rho_2\rho_1}$ for geodesics joining pure states} \label{sec-regularization_geodesics_pure_states}

In this appendix we determine numerically the contraction $R_{\rho_2\rho_1}$ appearing in the
formula (\ref{eq-Bures_geodesics_reg_bis}) of the regularized geodesics 
when $\rho_1= \ketbra{\psi_1}{\psi_1}$ and $\rho_2= \ketbra{\psi_2}{\psi_2}$ are orthogonal pure states. This contraction is defined 
by the limit (\ref{eq-def_R_rho2_rho1}). Our results give numerical support to the conjecture that for any contraction $R_{21}: S_{21} \to S_{12}$, there exist regularization curves of $\rho_1$ and $\rho_2$ such that $R_{\rho_2\rho_1}= R_{21}$.  
More precisely, 
 since for pure states $S_{12} = \complex \ket{\psi_2}$ and $S_{21} = \complex \ket{\psi_1}$, one has from 
(\ref{eq-def_R_rho2_rho1})
\begin{equation} \label{eq-contraction_R_for_2_pure_states}
	R_{\rho_2\rho_1} = z_{21} \ketbra{\psi_2}{\psi_1}\; , \, z_{21} = \lim_{\eps_1,\eps_2 \to 0} \bra{\psi_2} U_{21} (\veceps) \ket{\psi_1}\;,
\end{equation}
see Sec.\ref{sec-regularization_method}. Here, $U_{21} (\veceps)$ is the unitary in the polar decomposition of $\sqrt{\rho_2(\eps_2)} \sqrt{\rho_1(\eps_1)}$, which can be determined numerically using standard singular value decomposition algorithms. 
We give below a family of regularization curves
$\eps_i \mapsto \rho_i(\eps_i)$ depending on two parameters, such that arbitrary values of $z_{21}$ in the complex closed unit disc $\overline{D}_1$ are obtained  by varying these parameters.
Since $R_{\rho_2\rho_1}: S_{21}\to S_{12}$ is a contraction if and only if $|z_{21}|\leq 1$, this shows that the above conjecture is true for orthogonal pure states $\rho_1$ and $\rho_2$. 

Note that the analytical results of Sec.~\ref{sec-geodesics_qubit} show that the aforementioned conjecture is true for a qubit. 
For the sake of simplicity, we consider a qutrit system. Let 
$\{\ket{1},\ket{2},\ket{3}\}$ be an
orthonormal basis  of $\complex^3$ such that $\ket{1}= \ket{\psi_1}$ and $\ket{2}= \ket{\psi_2}$.
We introduce the family of regularization curves of $\rho_1$ and $\rho_2$ parametrized by a complex number $c = r e^{i\phi}$, with $0 \leq r \leq 1$ and $0 \leq \phi \leq 2 \pi$, defined by the following matrices in the basis $\{\ket{1},\ket{2},\ket{3}\}$
	\begin{align} \label{eq-family_regularization_curves}
		\nn
		\rho^{(c)}_{1}(\varepsilon_1)&=
		\begin{pmatrix}
			1-2\varepsilon_1 &
			(r-\varepsilon_1)e^{i\phi} \eta_1 & 0\\
			(r-\varepsilon_1)e^{-i\phi}\eta_1 &
			\varepsilon_1 & 0\\
			0 & 0 & \varepsilon_1
		\end{pmatrix},
		\\[1ex]
		\rho^{(c)}_{2}(\varepsilon_2)&=
		\begin{pmatrix}
			\varepsilon_2 &
			(r-\varepsilon_2)e^{i\phi}\eta_2 & 0\\
			(r-\varepsilon_2)e^{-i\phi}\eta_2 &
			1-2\varepsilon_2 & 0\\
			0 & 0 & \varepsilon_2
		\end{pmatrix},
	\end{align}
with $\eta_i = \sqrt{\varepsilon_i (1-2\varepsilon_i)}$.
Clearly $\rho^{(c)}_{i}(\varepsilon_i) \to\rho_{i}$ as $\varepsilon_i\to 0$, $\tr[\rho^{(c)}_{i}(\varepsilon_i)]=1$, and
both matrices are strictly positive for $0<\varepsilon_i<1/2$ and $0\le r\le 1$.
The corresponding value of $z_{21}$ in (\ref{eq-contraction_R_for_2_pure_states}) is computed numerically as a function of $r$ and $\phi$. 
To this end, we 
investigate the convergence of $z_{21}(\eps_1,\eps_2)= \bra{\psi_2} U_{21} (\eps_1,\eps_2) \ket{\psi_1}$ when
$\varepsilon_1=\varepsilon_2$ are very small,
for discrete values of $r$ and $\phi$,
\begin{equation} \label{eq-discrete_value_r_phi}
r_m = \sqrt{\frac{m}{N_{\rm{grid}}-1}}\quad , \quad
\phi_n = \frac{2\pi n}{N_{\rm{grid}}-1}\; ,
\end{equation}
with $m,n = 0, \dots, N_{\rm{grid}}-1$. 
We find numerically that $|z^{(c_{m,n})}_{21}(\eps,\eps)|$ converges to  $r_m$. Fixing a sufficiently small $\varepsilon$ (e.g. \(\varepsilon = 10^{-14}\)) ensuring convergence, we show   the values of  $z^{(c_{m,n})}_{21}(\eps,\eps)$ for all $n$ and $m$ in Fig.~\ref{fig:z21_plane}.
The results in this figure indicate that for all $c \in \complex$, $|c|\leq1$,
\begin{equation}
z^{(c)}_{21} = \lim_{\varepsilon\to0} z^{(c)}_{21}(\eps,\eps)
= -\overline{c}\; , 
\end{equation}
 thus
 providing strong numerical evidence that the map $c \in \overline{D}_1 \mapsto z^{(c)}_{21}$ covers the unit disk.
This support the aforementioned conjecture in the special case of two pure states $\rho_1$ and $\rho_2$.

\begin{figure}[htbp]
	\begin{center}
		\includegraphics[width=7cm,angle=0]{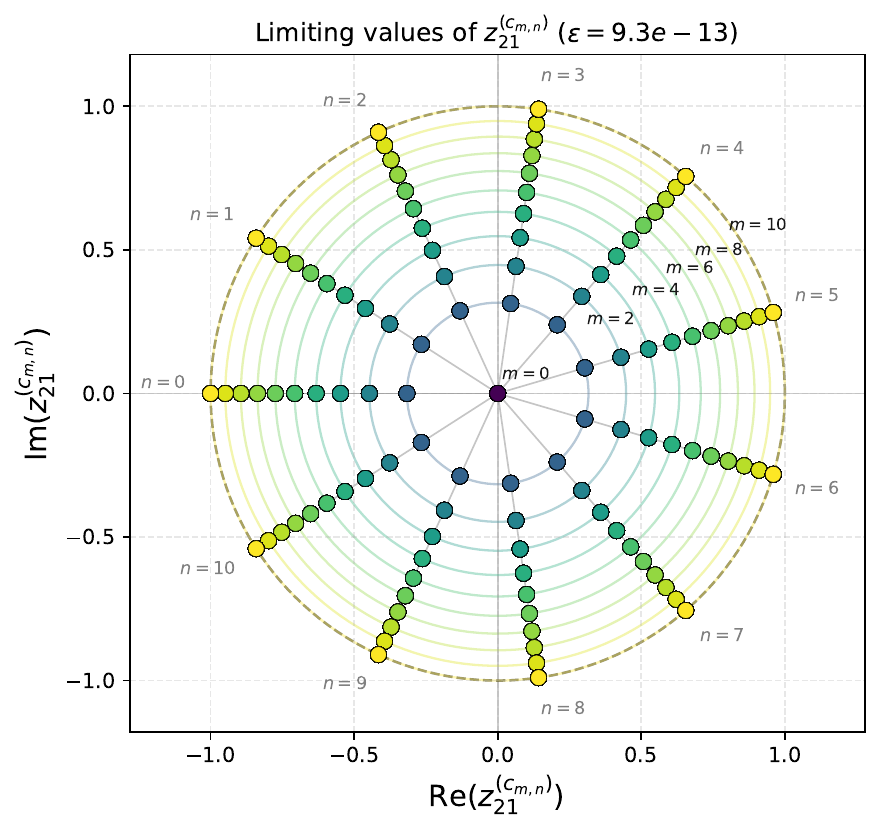}
	\end{center}  
	\captionsetup{format=plain,justification=raggedright}
	\caption{
		Complex-plane representation of the limit
		$\lim_{\eps\to 0} z^{(c_{m,n})}_{21}(\eps,\eps)$ for 
		the regularization curves (\ref{eq-family_regularization_curves})
	  with $r=r_m$ and $\phi=\phi_n$ as in (\ref{eq-discrete_value_r_phi}).
The colored circles and  gray rays indicate resp. the radial index $m$ and phase index $n$. The distribution of points shows that the limiting values fill the closed unit disk. The value $r=1$ leads to Fubini geodesics, given by (\ref{eq-geodesics_joining_pure_states_regularization}) with $|z_{21}|=1$ and arbitrary phases $\phi$; the values $0<r<1$ lead to elliptic rank-2 geodesics, and $r=0$ corresponds to the rank-2 straight diameter geodesic.}
	\label{fig:z21_plane}
\end{figure}

\section{Proof of Theorem~2} \label{sec-proof_theorem1}

In this appendix the proof of Theorem~2 is given. We determine the shortest geodesics joining two states $\rho_1$ and $\rho_2\in \partial \states$ satisfying
$\ker{\rho_1} \perp \ker{\rho_2}$ and contained in the interior of $\states$, save for a finite number of intersection points with the boundary.
Such geodesics  start from an invertible state $\rho_0\in \states^\inv$ and are specified by another state $\sigma$, which is taken to be $\rho_1$.
In fact, geodesics from an invertible state to a state on $\partial \states$ are well-defined and given  by (\ref{eq-Bures_geodesics_V}).

Let us first recall some known results about the intersection states of the geodesics with the boundary~\cite{Ericson05,moi_geodesics_QMetrology}. 
Eq.
(\ref{eq-Bures_geodesics_V}) can be written as
\begin{eqnarray} \label{eq-Bures_geodesics_invertible_states_bis}
\nn
\gamma_\geo (\tau) & = & X_0(\tau)\, \rho_0 \,X_0(\tau),
\\
X_0(\tau) & = & \frac{1}{\sin\tau_1} \big( \sin\tau\, M_0 + \sin (\tau_1-\tau)\identity\big)\;,
\end{eqnarray}  
where  $M_0$ is given by (\ref{eq-definition_M})
with $( \rho, \sigma) =(\rho_0, \rho_1)$ and $\tau_1 = d_{\Bures} ( \rho_0, \rho_1)$ is the first intersection time
of $\gamma_\geo$ with the boundary. 
The other intersection times $\tau_j$, $j=2,\ldots,q$, can be found from the condition 
$\det \gamma_\geo(\tau_j)= 0$. In view of (\ref{eq-Bures_geodesics_invertible_states_bis}) and  $\det  \rho_0 \not=0$, this condition reads $\det X_0(\tau_j)=0$. As $\det X_0(\tau)$ is the characteristic polynomial of $M_0$, this leads to an eigenvalue problem. Denoting by $\mu_1<\mu_2<\ldots < \mu_q$ the eigenvalues of $M_0$ in increasing order, one has 
\begin{equation} \label{eq-intersection_times}
\frac{\sin(\tau_1-\tau_j)}{\sin \tau_j} = - \mu_j\;,j=1,\ldots,q\;.
\end{equation} 
The first eigenvalue is thus $\mu_1=0$. 
Solving (\ref{eq-intersection_times}) for the $\tau_j$ one finds 
${\rm{cotan}}  (\tau_j) = (\cos \tau_1 - \mu_j)/\cos \tau_1$, hence showing that $0< \tau_1 < \tau_2 < \cdots < \tau_q$. Our assumption that $\gamma_\geo$ follows the shortest path from $\rho_1$ to $\rho_2$ 
imposes that $\rho_2$ is the second intersection state 
of $\gamma_\geo$ with the boundary. Thus consistency requires $\rho_2=\gamma_\geo (\tau_2)$.

\vspace{2mm}

\noindent {\bf Step 1: necessary and sufficient condition on $\rho_0$ such that $\rho_1,\rho_2 \in \gamma_\geo([0,\pi])$:} 

\vspace{1mm}

As  will be justified below, the intersection states $\rho_j = \gamma_\geo (\tau_j)$ have kernels given by the eigenspaces $P_j \Hh$ of $M_0$~\cite{Ericson05,moi_geodesics_QMetrology}. Hereafter, 
we write the spectral decomposition
\begin{equation} \label{eq-spectral_decomp_M}
M_0 = \sum_{j=1}^q \mu_j P_j\;.
\end{equation}
Therefore  
\begin{equation} \label{eq-P1_and_P2}
P_1 = \Pi_1^\bot\;,\;P_2= \Pi_2^\bot\;,\; \sum_{k=3}^q P_k = \Pi\;,
\end{equation}
where  $\Pi = \identity - \Pi_1^\bot - \Pi_2^\bot$
is the orthogonal projector onto $\Hh_1 \cap \Hh_2 = (\Hh_1^\bot \oplus \Hh_2^\bot)^\bot$.

The condition that the geodesic passes through $\rho_1$ and $\rho_2$ at times $\tau_1$ and $\tau_2$ ,
\begin{equation} \label{eq-condition_intersection_at_rho1rho2}
\gamma_\geo (\tau_i) = X_0(\tau_i)\rho_0 X_0(\tau_i)= \rho_i\;,\;i=1,2\;,
\end{equation}
can be rewritten thanks to (\ref{eq-Bures_geodesics_invertible_states_bis}),
(\ref{eq-intersection_times}) and (\ref{eq-spectral_decomp_M}) as
%
%
%
\begin{equation} \label{eq-condition_intersection_at_rho1rho2bis}
\frac{\sin^2 \tau_i}{\sin^2 \tau_1}
\sum_{j,k=1}^q (\mu_j-\mu_i)(\mu_k-\mu_i) P_j \rho_0 P_k = \rho_i\;,\; i =1,2\;. 
\end{equation}
Left- and right-multiplying (\ref{eq-condition_intersection_at_rho1rho2bis}) by $P_j$ and $P_k$ yields 
\begin{equation}\label{eq-condition_intersection_at_rho1rho2ter}
P_j \rho_0 P_k = \frac{\sin^2 \tau_1}{\sin^2 \tau_i} \frac{P_j \rho_i P_k}{(\mu_j-\mu_i)(\mu_k -\mu_i)}
\end{equation}
for any $j,k \in \{ 1,\ldots ,q\} \setminus \{i\}$.
Taking $j=k\not=i$ in (\ref{eq-condition_intersection_at_rho1rho2ter}) and using
the invertibility of $\rho_0$, one deduces that  
$P_j \rho_i P_j >0$ for any $j \not= i$, showing that 
$P_j \Hh \subseteq \supp  (\rho_i) $. 
Furthermore, (\ref{eq-condition_intersection_at_rho1rho2bis}) implies 
$P_i \rho_i P_i=0$ and thus $\ker \rho_i \subseteq P_i \Hh$.
This justifies  that $\ker ( \rho_i ) = P_i\Hh$, as claimed above.

Let us introduce the measurement probabilities
\begin{equation}
p_{j|i} = \tr ( P_j \rho_i)\;,\;1\leq j \leq q, i = 0,1,2\;.
\end{equation}
One has $p_{j|i}>0$ for any $j \not= i$ from the argument above,
and (\ref{eq-P1_and_P2}) entails $p_{i|i} = 0$ for $i=1,2$. 
This means that the projective measurement $\{ P_j\}_{j=1}^q$ implements an unambiguous discrimination scheme for discriminating $\rho_1$ and $\rho_2$. 
Since the kernels of $\rho_1$ and $\rho_2$ are orthogonal, one infers from  (\ref{eq-P1_and_P2}) that this measurement is optimal, \ie, it minimizes the probabilities $p_{? |i}= \sum_{k\geq 3} p_{k|i}$ of inconclusive outcomes~\cite{myreview}. 
Taking the trace of (\ref{eq-condition_intersection_at_rho1rho2ter}) for $j=k=1$ and $i=2$ and for  $j=k=2$ and $i=1$ one obtains
\begin{equation} \label{eq-relation_conditional_proba}
p_{1|0} = \frac{\sin^2 \tau_1}{\sin^2 \tau_2} \frac{p_{1|2}}{\mu_2^2}\;,\;
p_{2|0} = \frac{p_{2|1}}{\mu_2^2}\;,
\end{equation}
where we used $\mu_1=0$.

Let us introduce the post measurement conditional states associated with outcomes $1$ and $2$,
\begin{equation} \label{eq-rho_1|2}
\rho_{2|1} = \frac{\Pi_2^\bot \rho_1 \Pi_2^\bot}{p_{2|1}}
\;,\; \rho_{1|2} = \frac{\Pi_1^\bot \rho_2 \Pi_1^\bot}{p_{1|2}}\;.
\end{equation}
Eq. (\ref{eq-condition_intersection_at_rho1rho2ter}) is equivalent to the following set of equations: 
\begin{equation} \label{eq-equations_fixing_rho0}
\begin{array}{ccl}
\Pi_1^\bot \rho_0 \,\Pi_1^\bot & = & p_{1|0} \rho_{1|2}\;,
\\[1mm]
\Pi_2^\bot \rho_0 \,\Pi_2^\bot & = & p_{2|0} \rho_{2|1}\;,
\\[1mm]
\Pi_1^\bot \rho_0 \, \Pi & = &\dss- \frac{\sin^2\tau_1}{\mu_2 \sin^2\tau_2}\, \Pi_1^\bot \rho_2 (M_0-\mu_2)^{-1} \Pi\;,
\\[4mm]
\Pi_2^\bot \rho_0 \,\Pi & = & \Pi_2^\bot M_0^{-1} \rho_1 M_0^{-1} \Pi\;,
\\[1mm]
\Pi \, \rho_0 \,\Pi & = & \Pi \, M_0^{-1} \rho_1 M_0^{-1} \Pi \;,
\\[1mm]
\Pi \, \rho_0 \,\Pi
& =& 
\dss \frac{\sin^2 \tau_1}{\sin^2\tau_2} \Pi ( M_0 - \mu_2)^{-1} \rho_2 (M-\mu_2)^{-1} \Pi\;.
\end{array}
\end{equation}
These equations are, together with (\ref{eq-P1_and_P2}), equivalent to (\ref{eq-condition_intersection_at_rho1rho2}).
They fix the unknown state $\rho_0$ up to the off-diagonal 
blocks $\Pi_1^\bot \rho_0 \Pi_2^\bot$ and its adjoint $\Pi_2^\bot \rho_0 \Pi_1^\bot$, which are left arbitrary. The only constraint on these 
blocks comes from the fact that $\rho_0$ is an invertible density matrix.

Let us stress that the first intersection time $\tau_1$ can be fixed arbitrarily, as this amounts to moving $\rho_0$ along the geodesic (one should, however, choose $\tau_1$ sufficiently small to ensure that $\gamma_\geo$ does not hit the boundary between $\rho_0$ and $\rho_1$, so that it corresponds to the shortest geodesic arc $\rho_0 \to \rho_1$). 
 
 \vspace{2mm}
 
 \noindent {\bf Step 2: determination of the projectors $P_k$, $k \geq 3$:} 

\vspace{1mm}

The equality of the right-hand sides of the penultimate and ultimate equations in (\ref{eq-equations_fixing_rho0})
imposes extra conditions, from which one can determine the $\mu_2,\ldots,\mu_q$ and the projectors $P_3,\ldots, P_q$. Left- and right-multiplying these equations by $P_k$ and taking the trace, one finds
\begin{equation} \label{eq-equation_for_mu_k}
\frac{\mu_k-\mu_2}{\mu_k} = \frac{\sin \tau_1}{\sin \tau_2} \lambda_k\;,
\;k=3,\ldots, q\;,
\end{equation}
where we have set $\lambda_k = \sqrt{p_{k|2} /p_{k|1}}$.
An additional equation is needed apart from (\ref{eq-equation_for_mu_k})
in order to determine the $(q-1)$ eigenvalues $\mu_2,\ldots,\mu_q$. Such an equation is obtained by requiring $\tr \rho_0 =1$. Using (\ref{eq-equations_fixing_rho0}) this gives 
\begin{equation}\label{eq-additional_equation_for_mu_k}
\frac{\sin^2\tau_1}{\sin^2\tau_2} \frac{p_{1|2}}{\mu_2^2} + \frac{p_{2|1}}{\mu_2^2}+ \sum_{k=3}^q \frac{p_{k|1}}{\mu_k^2} = 1\;.
\end{equation}

Let us focus on finding the projectors $P_k$,
postponing the determination of $\mu_2,\ldots,\mu_q$ from (\ref{eq-equation_for_mu_k}) and (\ref{eq-additional_equation_for_mu_k}).  
Note that  $\Pi=0$ when 
$r_1+r_2=n$. Hence we may restrict ourselves to the case $r_1+r_2>n$, for which $q\geq 1$
 (in fact, $\Pi\Hh$ has dimension $r_1+r_2-n$).
 The equality of the penultimate and ultimate equations in 
(\ref{eq-equations_fixing_rho0})  reads
\begin{equation} \label{eq-conditions_on_P_k}
P_j \rho_2 P_k = \lambda_j \lambda_k P_j \rho_1 P_k\;,\;j,k=3,\ldots , q\;.
\end{equation}
%
Let us introduce the non-negative operator
\begin{equation} \label{eq-definition_M_12}
M_{12} = \sum_{k=3}^q \lambda_k P_k\;.
\end{equation}
Then (\ref{eq-conditions_on_P_k}) is equivalent to
\begin{equation}
\Pi \rho_2 \Pi = M_{12} \rho_1 M_{12}\;.
\end{equation}
Left- and right-multiplying by $\sqrt{\rho_1}$, one gets
\begin{equation} \label{eq-equation_giving_M_12}
\sqrt{\rho_1 }\,\Pi \rho_2 \Pi \sqrt{\rho_1}= (\sqrt{\rho_1 } M_{12} \sqrt{\rho_1})^2\;.
\end{equation}
But
\begin{equation} \label{eq-Pi_sandwitched_between_rho_2_and_rho_1}
\sqrt{\rho_2} \,\Pi \sqrt{\rho_1} = \sqrt{\rho_2} ( \identity - \Pi_1^\bot - \Pi_2^\bot )\sqrt{\rho_2} =  \sqrt{\rho_2}\sqrt{\rho_1}\;.
\end{equation}
%
Plugging this relation in (\ref{eq-equation_giving_M_12}) and taking the square root (observe that $\sqrt{\rho_1} M_{12} \sqrt{\rho_1} \geq 0$ as $M_{12} \geq 0$), one finds $|\sqrt{\rho_2}\sqrt{\rho_1}|= \sqrt{\rho_1} M_{12} \sqrt{\rho_1}$, \ie,
\begin{equation} \label{eq-expressin_M_12}
M_{12} = \rho_1^{-1/2} \Lambda_{\rho_2\rho_1} \rho_1^{-1/2} = M_{\rho_1\rho_2}\;,
\end{equation}
where we have used $M_{12} = \Pi M_{12} \Pi$ and $\Pi \Hh \subset \Hh_1$. 
Comparing with (\ref{eq-spectral_decomp_M}) this shows that the projectors $P_k$, $3 \leq k \leq q$, are the spectral projectors of $M_{\rho_1\rho_2}$ associated to its nonzero eigenvalues $\lambda_k$. Conversely, if this holds then 
(\ref{eq-conditions_on_P_k}) is satisfied.

An explicit calculation using $P_k = P_k \rho_1^{-1/2} \sqrt{\rho_1}$ and $\sqrt{\rho_1}\sqrt{\rho_2} U_{\rho_2\rho_1} = \Lambda_{\rho_2\rho_1}$
shows that
the spectral projectors  $P_k$ and eigenvalues $\lambda_k$ of $M_{\rho_1\rho_2}$ satisfy the identity
\begin{equation} \label{eq-necessary_and_sufficient_cond_optimal_meas}
P_k \sqrt{\rho_2} \,U_{\rho_2\rho_1} = \lambda_k P_k \sqrt{\rho_1} \;,\;k=3,\ldots,q\;.
\end{equation}
It is known that (\ref{eq-necessary_and_sufficient_cond_optimal_meas}) is a necessary and sufficient condition for the measurement $\{ P_j\}$ to maximize the 
classical Hellinger distance between the two outcome probabilities
$\{ p_{j|1}\}$ and $\{ p_{j|2}\}$ (see e.g.~\cite{myreview}, Proposition 7.C.1 and its proof). For such a measurement, the classical Hellinger distance is equal to the Bures distance $d_{\Bures}(\rho_1,\rho_2)$, or  equivalently, the bound~\cite{Nielsen} 
\begin{equation}
\sqrt{F(\rho_1,\rho_2)} \leq \sum_{j=1}^q \sqrt{p_{j|1} p_{j|2}}
\end{equation}
becomes an equality. This means that the measurement $\{ P_j\}_{j=1}^q$ is not only optimal for unambiguously discriminating $\rho_1$ and $\rho_2$, but is also optimal for distinguishing the two states from the statistics of the measurement data.
Since $p_{i|i}=0$ for $i=1,2$, one has
\begin{equation} \label{eq-fidelity_in_terms_of_proba}
\sqrt{F(\rho_1,\rho_2)} = \sum_{j=3}^q \sqrt{p_{j|1} p_{j|2}}\;.
\end{equation}

\vspace{1mm}
  
\noindent {\bf Step 3: Determination of the eigenvalues $\mu_j$.}

\vspace{1mm}
 
We now proceed to determine $\mu_2,\ldots,\mu_q$ and the intersection times $\tau_2,\ldots,\tau_q$ from (\ref{eq-intersection_times}), (\ref{eq-equation_for_mu_k}) and (\ref{eq-additional_equation_for_mu_k}).
Note that $\tau_2$ should be given by
 $\tau_2-\tau_1 = d_{\Bures} (\rho_2,\rho_1)$ because by hypothesis 
$\gamma_\geo|_{[\tau_1,\tau_2]}$ is a shortest geodesic arc $\rho_1\to \rho_2$. 
Since one has some freedom in choosing $\tau_1$, it is convenient to take
$0 < \tau_1 \ll 1$, \ie, the invertible state $\rho_0$ is very close to $\rho_1$.
Then, expanding the \LHS of (\ref{eq-intersection_times}), one has
\begin{equation} \label{eq-relation_between_mu_j_and_tau_j}
\mu_j = 1 - \tau_1 \cotan (\tau_j) + O (\tau_1^2)\;,\;j=2,\ldots,q\;.
\end{equation}
Substituting in (\ref{eq-equation_for_mu_k}), one finds
\begin{equation} \label{eq-equation_for_mu_k_bis}
\cotan ( \tau_2) - \cotan (\tau_k)= \frac{\lambda_k}{\sin \tau_2} + O (\tau_1)\;,\;k=3,\ldots,q\;.
\end{equation}
Similarly, one may substitute (\ref{eq-relation_between_mu_j_and_tau_j}) into 
 (\ref{eq-additional_equation_for_mu_k}), expand in powers of $\tau_1$  and identify terms of order $0$ and of order $1$ to find
\begin{equation}
\left\{
\begin{array}{l}
\dss p_{2|1} + \sum_{k=3}^q p_{k|1}  =  1
\\
\dss p_{2|1} \cotan \tau_2 + \sum_{k=3}^q p_{k|1} \,\cotan \tau_k = 0
\end{array}
.
\right.
\end{equation} 
 The first equation is trivially satisfied since $p_{1|1}=0$ and $\sum_{j=1}^q p_{j|1}=1$. Plugging (\ref{eq-equation_for_mu_k_bis}) into the second equation and rearranging terms, one obtains
 \begin{equation}
 \cos \tau_2 = \sum_{k=3}^q \lambda_k p_{k|1 }+O(\tau_1)= \sum_{k=3}^q \sqrt{ p_{k|1} p_{k|2}} + O(\tau_1)\;.
 \end{equation}
Thanks to (\ref{eq-fidelity_in_terms_of_proba})
the \RHS  is nothing but the square root of the fidelity $F(\rho_1,\rho_2)$. Hence $\tau_2 = d_{\rm{B}} (\rho_1,\rho_2)+ O(\tau_1)$, in agreement with the anticipated result. 
The other intersection times are given to lowest order in $\tau_1$ by
\begin{equation}
\cotan \tau_k = \sqrt{1-F(\rho_1,\rho_2)} \big( \sqrt{F(\rho_1,\rho_2)} - \lambda_k \big) \;.
\end{equation}

\vspace{1mm}

\noindent {\bf Step 4: Determination of the geodesics $\gamma_\geo$.}

\vspace{1mm}

We can now determine the geodesics explicitly.  
Let us first note that $\rho_0$ is 
at distance $\tau_1$ from $\rho_1$, so that 
$\rho_0 =\rho_1 + O(\tau_1)$ for small $\tau_1$. Thus
$\Pi_1^\bot \rho_0 \Pi_2^\bot$ and its adjoint 
must be of order one in $\tau_1$. We set
\begin{equation} \label{eq-def_dot_rho_12}
\Pi_1^\bot \rho_0 \Pi_2^\bot + \Pi_{2}^\bot \rho_0 \Pi_1^\bot = 
- \tau_1 \dot{\rho}_1^{(12)} + O(\tau_1^2)\;.
\end{equation}
Recall that the off-diagonal term
$\Pi_1^\bot \rho_0 \Pi_2^\bot$ is not fixed by 
(\ref{eq-equations_fixing_rho0}). Hence
 $\dot{\rho}_1^{(12)}$ is an arbitrary self-adjoint operator with kernel $\Pi \Hh$ satisfying  
$\Pi_i^\bot \dot{\rho}_1^{(12)} \Pi_i^\bot =0$ for $i=1,2$. 
Using (\ref{eq-relation_between_mu_j_and_tau_j}) and (\ref{eq-equation_for_mu_k_bis}) and expanding in powers of $\tau_1$
the operators $M_0(\tau)$ and $X_0(\tau)$ given by (\ref{eq-Bures_geodesics_invertible_states_bis}) and (\ref{eq-spectral_decomp_M}),
one finds
\begin{align} \label{eq-expansion_M_0}
&  M_0   =  \Pi_1 - \frac{\tau_1}{\sin \tau_2} \Big( \cos\tau_2 \, \Pi_1  - M_{12} \Big) 
+ O(\tau_1^2)\;,
\\ \label{eq-expansion_X_0}
&  X_0(\tau)  =  
\frac{1}{\sin \tau_1} \Big( - \sin \tau \,\Pi_1^\bot 
+ \tau_1  X_{12} ( \tau)
\\ \nn
&  \hspace{1.5cm}  + \tau_1 \cotan\tau_2 \sin \tau \,\Pi_1^\bot \Big)  + O ( \tau_1^2)\Big)\;,
\\
\label{eq-expansion_inverse_M_0-mu_2}
& (M_0 - \mu_2)^{-1} \Pi  =  \frac{\sin \tau_2}{\tau_1} M_{12}^{-1} \Pi + O (1)
\;,
\end{align}
where we have used $\Pi_1 = \Pi_2^\bot + \Pi$ and have set 
\begin{equation}
X_{12} ( \tau) = \frac{1}{\sin \tau_2} \Big( \sin \tau \,M_{12} + \sin (\tau_2 - \tau)\identity \Big)\;.
\end{equation}
One deduces from (\ref{eq-relation_conditional_proba}), (\ref{eq-rho_1|2}) and (\ref{eq-equations_fixing_rho0}) that
\begin{eqnarray}
\nn
\rho_0 & = & \Big( \frac{\sin\tau_1}{\mu_2 \sin \tau_2}\Big)^2 \Pi_1^\bot \rho_2 \Pi_1^\bot
 + \frac{1}{\mu_2^2} \Pi_2^\bot \rho_1 \Pi_2^\bot - \tau_1 \dot{\rho}_1^{(12)}
\\ 
& & 
+ \frac{1}{\mu_2} \Big( -\Big( \frac{\sin \tau_1}{\sin \tau_2} \Big)^2 \Pi_1^\bot \rho_2 (M_0-\mu_2)^{-1} \Pi 
\\ \nn
& &
+ \Pi_2^\bot \rho_1 M_0^{-1} \Pi
+ {\rm{h.c.}}
\Big)
+ \Pi M_0^{-1} \rho_1 M_0^{-1} \Pi\;.
\end{eqnarray}  
Replacing this expression into (\ref{eq-Bures_geodesics_invertible_states_bis}) and using the relation
\begin{equation}
M_{12} \rho_1 M_{12} = M_{\rho_2\rho_1}\rho_1 M_{\rho_2\rho_1} 
= \Pi_1 \rho_2 \Pi_1 = \Pi \rho_2\Pi\;, 
\end{equation}
one finds to lowest order in $\tau_1$
%
\begin{eqnarray} 
\nonumber
&& 
\gamma_{\geo} ( \tau) =\frac{1}{\sin^2 \tau_2}
\bigg( \sin^2 ( \tau_2 - \tau)\,  \rho_1  + \sin^2 ( \tau )\, \rho_2
\\ \nn
& & 
+\sin ( \tau_2 - \tau)  \sin ( \tau) \Big( \Big(
\rho_1 M_{12} + \Pi_1^\bot \rho_2 M_{12}^{-1} \Pi 
 + {\mathrm{h.c.}} \Big)
 \\
& &   
+\sin \tau_2 \, \dot{\rho}^{(12)}_1 
\Big)
\bigg) + O(\tau_1)\;.
\end{eqnarray}
%
Using the notation (\ref{eq-definition_S_12})-(\ref{eq-decomp_Pi_i}) in the main text, one has 
$\supp ( \Lambda_{\rho_1\rho_2}) = \sqrt{\rho_2} \Hh_1 \supseteq \sqrt{\rho_2}
\, \Hh_1 \cap \,\Hh_2$, see (\ref{eq-support_M_rho1_rho2}).
(Note that the reverse inclusion also holds when $\Hh_1^\bot \perp \Hh_2^\bot$ 
by virtue of (\ref{eq-support_M_rho1_rho2}) 
and $\Lambda_{\rho_1\rho_2} = \sqrt{\rho_2}\, M_{\rho_2\rho_1} \sqrt{\rho_2}$). Thus
%
\begin{equation} \label{eq-relation_with_projector_Pi}
\sqrt{\rho_2}\,\Pi = \Pi_{\rho_1\rho_2} \sqrt{\rho_2} \,\Pi\; ,
\end{equation}
%
with a similar relation by exchanging $1$ and $2$.
One has from (\ref{eq-expressin_M_12}) and (\ref{eq-polar_decomp_non_invertible_states})
\begin{eqnarray} \label{eq-rho_M}
\nn
\rho_1 M_{12} & = & 
\sqrt{\rho_1} \Lambda_{\rho_2\rho_1} \rho_1^{-1/2} \Pi
\\
& = & \sqrt{\rho_1} \big( U_{\rho_2\rho_1}^{(0)}\big)^\dagger \sqrt{\rho_2} \,\Pi\;. 
\end{eqnarray}
Note that since $U_{\rho_2\rho_1}^{(0)\,\dagger}$ is uniquely defined as an operator from
$\Pi_{\rho_1\rho_2}\Hh$ to $\Pi_{\rho_2\rho_1} \Hh$, thanks to 
(\ref{eq-relation_with_projector_Pi}) the \RHS of (\ref{eq-rho_M}) is well-defined
and has values in $\rho_1 \Hh_2$. 

Similarly, 
\begin{eqnarray} \label{eq-rho_M^-1}
\nn
\Pi_1^\bot \rho_2 M_{12}^{-1} \Pi 
& = & \Pi_1^\bot \rho_2 \Pi M_{12}^{-1}
\\ \nn
& = & 
\Pi_1^\bot \rho_2 \Pi \sqrt{\rho_1} \,\Lambda_{\rho_2\rho_1}^{-1} \sqrt{\rho_1}
\\ \nn
& = & \Pi_1^\bot \rho_2 \sqrt{\rho_1} \,\Lambda_{\rho_2\rho_1}^{-1} \sqrt{\rho_1}
\\
& = & 
\Pi_1^\bot \sqrt{\rho_2}\, U_{\rho_2\rho_1}^{(0)} \sqrt{\rho_1} \;,
\end{eqnarray}
where we have used (\ref{eq-Pi_sandwitched_between_rho_2_and_rho_1}) in the third line and
 $\sqrt{\rho_2}\sqrt{\rho_1} \Lambda_{\rho_2\rho_1}^{-1} = U_{\rho_2\rho_1}^{(0)}$
 in the fourth line, with 
 $U_{\rho_2\rho_1}^{(0)}$ the uniquely-defined part of the unitary in the polar decomposition of $\sqrt{\rho_2}\sqrt{\rho_1}$, given by (\ref{eq-definition_U_21^(0)}).   
From (\ref{eq-rho_M}), (\ref{eq-rho_M^-1}) and $\Pi + \Pi_1^\bot = \Pi_2$ one infers
\begin{equation} \label{eq-formula_to_check}
M_{12}\, \rho_1 + \Pi_1^\bot \rho_2 M_{12}^{-1} \Pi  = \sqrt{\rho_2}\, U_{\rho_2\rho_1}^{(0)}
\sqrt{\rho_1}\;.
\end{equation}
%
Collecting the above results one finds (\ref{eq-Bures_geodesics_second_method}).

Changing $\dot{\rho}^{(12)}_1$ amounts to changing the tangent vector  of $\gamma_\geo^{1\to2}$ at  $\rho_1$,
\begin{eqnarray} \label{eq-tangent_vector}
\nn
\dot{\gamma}_\geo (0)& = &  \frac{1}{\sin \tau_2} \Big( 
\big( \sqrt{\rho_2} \,U_{\rho_2\rho_1}^{(0)} \sqrt{\rho_1} 
+ {\mathrm{h.c.}} \big)  + \sin \tau_2 \,\dot{\rho}^{(12)}_1
\\
& & - 2 \cos \tau_2 \rho_1 
\Big)\;,
\end{eqnarray}
therefore one has infinitely many different geodesics. 
To see which constraint on  $\dot{\rho}^{(12)}_1$ is set by the condition that  
$\rho_0$ is an invertible density matrix, \ie, $\rho_0 >0$, 
we write $\rho_0$ as a $3 \times 3$ block matrix with respect to the 
space decomposition $\Hh = \Hh_1^\bot \oplus \Hh_2^\bot \oplus \Pi^\bot \Hh$. 
 Thanks to 
(\ref{eq-relation_conditional_proba})-(\ref{eq-equations_fixing_rho0}) and (\ref{eq-def_dot_rho_12}), this matrix reads
\begin{widetext}
	\begin{equation} \label{eq-non_negativity_of_rho0}
{\mathrm{Mat}} (\rho_0 ) = 
\begin{pmatrix}
 \dfrac{\sin^2 \tau_1}{\mu_2^2 \sin^2 \tau_2}\Pi_1^\bot \rho_2 \Pi_1^\bot & \;\;
 - \tau_1 \Pi_1^\bot \dot{\rho}^{(12)}_1 \Pi_2^\bot 
 & \;- \dfrac{\sin^2 \tau_1}{\mu_2 \sin^2 \tau_2} \Pi_1^\bot \rho_2 ( M_0 - \mu_2)^{-1} \Pi
 \\[3ex]{eq-CNS_rho_0>0}
 - \tau_1 \Pi_2^\bot \dot{\rho}^{(12)}_1 \Pi_1^\bot  & \;\;
 \dfrac{1}{\mu_2^2} \Pi_2^\bot \rho_1 \Pi_2^\bot &\; 
 \Pi_2^\bot M_0^{-1} \rho_1 M_0^{-1} \Pi
 \\
 - \dfrac{\sin^2 \tau_1}{\mu_2 \sin^2 \tau_2} \Pi ( M_0 - \mu_2)^{-1} \rho_2  \Pi_1^\bot   &  \;\;
 \Pi M_0^{-1} \rho_1 M_0^{-1} \Pi_2^\bot & \;
 \Pi M_0^{-1} \rho_1 M_0^{-1} \Pi
\end{pmatrix}
\;>\; 0\;.
\end{equation}	
For some given $m \times m$ and $p \times p$ self-adjoint matrices $A$ and $D$ 
and a $m \times p$ rectangular matrix $B$, a necessary and sufficient condition for $ = 
\begin{pmatrix}
A & B \\
B^\dagger & D 
\end{pmatrix}
$
to be positive is  
\begin{equation} \label{eq-Schur_complement_CNS_positivity}
D>0 \;,\; A - B D^{-1} B^\dagger > 0\;,
\end{equation} 
where the last matrix is the Schur complement.
Applying this fact with $A$ the $2 \times 2$ upper left block and $D$ the $1 \times 1$ lower right block in (\ref{eq-non_negativity_of_rho0}) and using the 
expansions (\ref{eq-relation_between_mu_j_and_tau_j}), (\ref{eq-expansion_M_0}) and (\ref{eq-expansion_inverse_M_0-mu_2}) and the relation (\ref{eq-rho_M^-1}), one finds that $\rho_0 >0$ if and only if $\Pi \rho_1\Pi + O(\tau_1) >0$ and
the following matrix in $\Hh_1^\bot \oplus \Hh_2^\bot$ is positive
\begin{equation} \label{eq-CNS_rho_0>0}
\begin{pmatrix}
\dfrac{\tau_1^2}{\sin^2 \tau_2} \,\Pi_1^\bot \sqrt{\rho_2} \Big( \identity - \sqrt{\rho_2} ( \Pi \rho_2 \Pi)^{-1} \sqrt{\rho_2} \Big) \sqrt{\rho_2} \,\Pi_1^\bot
\!\! + \! O (\tau_1^3) & \!\!
- \dfrac{\tau_1}{\sin \tau_2} \Pi_1^\bot \Big(  \dot{\rho}^{(12)}_1- \sqrt{\rho_2}\, U_{\rho_2\rho_1}^{(0)} \sqrt{\rho_1} ( \Pi \rho_1 \Pi)^{-1} \rho_1 \Big)\Pi_2^\bot \!\! + \! O (\tau_1^2)
\\[3ex]
- \dfrac{\tau_1}{\sin \tau_2} \Pi_2^\bot \Big(  \dot{\rho}^{(12)}_1- \rho_1  
 ( \Pi \rho_1 \Pi)^{-1} \sqrt{\rho_1} \, U_{\rho_2\rho_1}^{(0)\;\dagger} \sqrt{\rho_2} \Big)\Pi_1^\bot \!\! + \! O (\tau_1^2)
& \!\!
\Pi_2^\bot \sqrt{\rho_1} \Big( \identity - \sqrt{\rho_1} ( \Pi \rho_1 \Pi)^{-1} \sqrt{\rho_1} \Big) \sqrt{\rho_1} \, \Pi_2^\bot \!\!+ \! O (\tau_1)
\end{pmatrix}
.
\end{equation}
\end{widetext}

\newpage



\end{document}